\documentclass[journal]{IEEEtran}
\usepackage{cite}

\ifCLASSINFOpdf
   \usepackage[pdftex]{graphicx}
  \graphicspath{{../pdf/}{../jpeg/}}
   \DeclareGraphicsExtensions{.pdf,.jpeg,.png}
\else
   \usepackage[dvips]{graphicx}
  \graphicspath{{../eps/}}
  \DeclareGraphicsExtensions{.eps}
\fi
\usepackage{amsmath,amssymb,amsfonts}

\usepackage{algorithmic}

\usepackage{textcomp}
\usepackage{xcolor}
\usepackage{hyperref}
\usepackage{subcaption}

\usepackage{array}
\begin{document}
%
\title{EXT-TAURUM P2T: an Extended Secure CAN-FD Architecture for Road Vehicles}
%
%
%

\author{Franco Oberti,~\IEEEmembership{Member,~IEEE,}
        Alessandro Savino,~\IEEEmembership{Member,~IEEE,}
        Ernesto Sanchez,~\IEEEmembership{Senior Member,~IEEE}
        Filippo Parisi,
        Stefano Di Carlo,~\IEEEmembership{Senior Member,~IEEE}
\thanks{F. Oberti, A. Savino, E. Sanchez and S. Di Carlo are with the Control and computer Engineering Department of Politecnico di Torino, Italy. Contact e-mail: (stefano.dicarlo@polito.it).}
\thanks{F. Oberti and F. Parisi are with PUNCH Torino S.p.A., Torino, Italy.}
\thanks{Manuscript received April 19, 2005; revised August 26, 2015.}}

\maketitle

\begin{abstract}
The automobile industry is no longer relying on pure mechanical systems; instead, it benefits from advanced Electronic Control Units (ECUs) in order to provide new and complex functionalities in the effort to move toward fully connected cars. However, connected cars provide a dangerous playground for hackers. Vehicles are becoming increasingly vulnerable to cyber attacks as they come equipped with more connected features and control systems. This situation may expose strategic assets in the automotive value chain. In this scenario, the Controller Area Network (CAN) is the most widely used communication protocol in the automotive domain. However, this protocol lacks encryption and authentication. Consequently, any malicious/hijacked node can cause catastrophic accidents and financial loss. Starting from the analysis of the vulnerability connected to the CAN communication protocol in the automotive domain, this paper proposes EXT-TAURUM P2T a new low-cost secure CAN-FD architecture for the automotive domain implementing secure communication among ECUs, a novel key provisioning strategy, intelligent throughput management, and hardware signature mechanisms. The proposed architecture has been implemented, resorting to a commercial Multi-Protocol Vehicle Interface module, and the obtained results experimentally demonstrate the approach's feasibility. 
\end{abstract}


%
\IEEEpeerreviewmaketitle

\section{Introduction}
Nowadays, our cars are providing a dangerous playground for hackers. Vehicles are becoming increasingly vulnerable to cyber attacks as they come equipped with connected features and control systems. This situation may expose strategic assets in the automotive value chain. The World Forum for Harmonization of Vehicle Regulations (WP.29), a working party of the UN Economic Commission for Europe (UNECE), confirms this trend~\cite{WP29}. WP.29 recently introduced the new UN R155~\cite{unece-155-2021} and UN R156~\cite{unece-156-2021} UNECE regulations for cyber-security in the automotive domain. These regulations explicitly mention four disciplines:
\begin{itemize}
\item \emph{Managing cyber-risks for vehicles}: each company shall set a security office in its organization for managing product security according to legislation guidelines. Cyber-security processes oversee risk assessment analysis, requirements, specifications, and incident reports.
    \item \emph{Securing vehicles ``by design" to mitigate risks along the value chain}: attacks are constantly evolving, increasing the number of threats and their level of risk. Security knowledge, competencies, capabilities, and new solutions are the mission to pursue by pushing increasingly secure products on the market.
    \item \emph{Detecting and responding to security incidents across vehicle fleets}:
     standard security regulations must be applied actively over the entire product lifetime. This translates into constantly monitoring the security of the vehicle's fleet and releasing incident reports for effective vehicle attacks.  
    \item \emph{Managing safe and secure updates of the vehicle software, including a legal basis for over-the-air updates}:
    companies shall create a framework to guarantee security not only considering single assets (i.e., product, IT services) but also considering the entire ecosystem with its infrastructure as a whole to minimize any risks concerning sensible data breaches.
\end{itemize}

UN R155 and UN R156 are non-negotiable and represent mandatory conditions for approval and market access to the entire UNECE WP.29 member countries with the addition of Japan and Korea. These regulations will be compulsory for new permanently and seamlessly connected vehicles from July 2022 and extended to existing cars by July 2024. 

These requirements create a challenging scenario for the entire automotive sector, requiring an effort to implement new cyber-security monitoring, detection, reporting, and response capabilities across the whole vehicle life-cycle with the involvement of the entire supply chain. Failing in fulfilling the UN R155 and UN R156 requirements means a production roadblock with a considerable loss of money. 

The Controller Area Network (CAN) is the most widely used communication protocol in the automotive domain. The CAN protocol was designed to guarantee reliable communication between electronic modules in high-noise environments. However, it lacks encryption and authentication. Consequently,  any malicious/hijacked node can cause catastrophic accidents and financial loss  \cite{bozdal2020evaluation}. Limited throughput and secret key availability are among the main limitations to implementing security mechanisms that guarantee the authenticity and integrity of CAN communications, thus posing strong constraints on the future development of permanently and seamlessly connected road vehicles.

This article reviews specific security vulnerabilities connected to the CAN Flexible Data rate (CAN-FD) architecture employed in road vehicles \cite{6884472}. It proposes a solution named Extended TAURUM P2T (EXT-TAURUM P2T) that increases the current security level in road vehicles by addressing the identified vulnerabilities and remaining strictly compliant with the UNECE regulations. The new approach, i.e., EXT-TAURUM P2T, extends the TAURUM P2T architecture presented in \cite{oberti2021taurum}, guaranteeing all the functionalities available in the previous architecture: 
\begin{itemize}
\item increased security with limited cost and hardware resources;
\item implementation of a rolling secret key system;
\item privilege separation;
\item secret key auto-generation without external key infrastructures.
\item throughput optimization for secure mechanism.
\item physical attack mitigation solution.
\end{itemize}

These functionalities are complemented by two novel features that represent the key novelty of the Extended TAURUM P2T architecture: 
\begin{itemize}
\item A new speculative MAC calculation functionality implemented on top of an OSEK operating system that enables to increase the capability of the system to support overloading situations that might be associated with DoS attacks;
\item The implementation of a hardware signature infrastructure that exploits the EXT-TAURUM P2T Secure CAN network to address a new hardware replacement attack. The high-level idea of this concept was initially introduced in \cite{9525579}, and this paper shows that the EXT-TAURUM P2T infrastructure provides all security primitives and facilities to move from the concept to actual implementation.
\end{itemize}

Overall, EXT-TAURUM P2T contributes to securing vehicles ``by design" by building a new onboard secure communication network providing the necessary security primitives. Sensible information can be securely exchanged among the different Electronic Control Units governing the vehicle's activities. 
Moreover, EXT-TAURUM P2T contributes to create a framework to guarantee security, considering the entire security infrastructure. The EXT-TAURUM P2T self key provisioning architecture removes the need for centralized key provisioning infrastructures. This scheme simplifies the carmakers' information management systems, making them easier to manage and reducing the connected risks.

The paper is organized as follows: \autoref{sec:background} introduces basic CAN network definitions required to understand the proposed techniques, while \autoref{sec:CAN-vulnerabilities} discusses the main vulnerabilities of this type of network. Section~\ref{sec:taurum-P2T} overviews EXT-TAURUM P2T secure architecture while  \autoref{sec:results} provides experimental results. Finally, \autoref{sec:conclusions} summarizes the main contributions and concludes the paper.

\section{Background}
\label{sec:background}

\subsection{Automotive CAN Network overview }
\label{sec:CAN-overview}

The automobile industry no longer relies on pure mechanical systems; instead, it benefits from many intelligent features based on advanced Electronic Control Units (ECUs) \cite{bozdal2020evaluation}. In a modern car, it is common to integrate more than 70 ECUs controlling various physical subsystems \cite{albert2004comparison}.


Communication is an essential element of this complex infrastructure since the different subsystems must control actuators or receive feedback from other subsystems.

The Controller Area Network (CAN) is the most widely used communication protocol for different interconnecting ECUs specified in the ISO 11898-1 standard. It is based on a flexible multi-cast serial bus that supports a software implementation of a wide range of safety, security, and convenience features. This flexibility reduces costs and complexity associated with ``hard-wired" solutions. The CAN Flexible Data rate (CAN-FD) is an extension to the original CAN bus protocol introduced by BOSCH \cite{6884472} to meet the need to increase the data transfer rate up to 5 times while enabling a significant increase in the message size to be used in modern cars.

The messages transmitted over the vehicle's CAN network have heterogeneous requirements in terms of accessibility (i.e., visibility outside the car) and security (i.e., confidentiality, integrity, and authenticity). In a standard automotive CAN network, several classes of messages must be accessible by external inspectors to satisfy specific country-based legal vehicle regulations (e.g., emission legislation). Consequently, these messages are transmitted in clear text, and vehicles are equipped with an On-Board Diagnostic (OBD) port to monitor the CAN network traffic. 

Therefore, when looking at the CIA triad's security pillars \cite{CIAtriad}, only the integrity and authenticity of CAN data frames can be implemented. This is achieved by reserving a portion of the CAN data frame to store a Cipher-based Message Authentication Code (CMAC) signature of the transmitted payload \cite{CMAC,MAC}. For preventing replay attacks \cite{Replyattack}, a rolling counter is usually included in each transmitted frame \cite{rollingcounter}.

\subsection{Automotive control modules overview }
\label{sec:ACM_overview}

In the automotive domain, ECUs are classified into three categories:
\begin{enumerate}
  \item \emph{hard-real-time}, performing highly safety-critical tasks,
  \item \emph{soft-real-time}, for mixed-critical functionalities, and
  \item \emph{non-real-time}, for performing the remaining tasks.
\end{enumerate}

Automotive safety-critical systems (i.e., hard- and soft-real-time) adopt specific real-time operating systems compliant to the OSEK (Offene Systeme und deren Schnittstellen für die Elektronik in Kraftfahrzeugen; English: "Open Systems and their Interfaces for the Electronics in Motor Vehicles") open standard \cite{osek}. OSEK was founded by a German automotive company consortium supported by the Karlsruhe Institute of Technology and included specifications for an embedded operating system (OS), a communications stack (COM), and a network management protocol (NM) for automotive embedded systems.

The OSEK specifications impact the embedded software architecture executed on an ECU. Applications are organized into ``tasks" that are statically defined at compile time with a fixed priority. Every task can assume three execution states: \texttt{SUSPENDED}, \texttt{READY}, and \texttt{RUNNING}. \texttt{READY} tasks are scheduled according to their priority. First In First Out (FIFO) scheduling is used for tasks with equal priority (i.e., round-robin scheduling is not permitted). When scheduled, a basic task runs to completion except when a higher priority task preempts it or an interrupt is detected. To make sure that real-time deadlines can be guaranteed, deadlocks and priority inversion are prevented by a  priority ceiling algorithm \cite{kluge2009implementing}. 

In most OSEK implementations, there is a zero-priority (i.e., low priority) idle task, also known as the background task. The ECU executes this task until an interrupt moves a different task from \texttt{SUSPENDED} to \texttt{READY}. The background task can be exploited to perform important activities such as:
\begin{itemize}
\item Idle time monitoring;
\item Low power microprocessor management;
\item Watchdog tickling;
\item Non-real-time custom activities;
\item Future extensions.
\end{itemize}

EXT-TAURUM P2T architecture exploits the custom activities provided by OSEK OS to optimize the computational effort to guarantee security as described in \autoref{sec:speMac}.

\section{Automotive CAN Network Vulnerabilities}
\label{sec:CAN-vulnerabilities}

In the automotive domain, two main categories of attackers may exploit CAN vulnerabilities to violate the vehicle's ECUs:
\begin{itemize}
    \item \emph{Vehicle owner}: not interested in damaging its good. Its goal is to improve vehicle performance or tamper with annoying features (e.g., diagnostic). 
    \item \emph{Professional attacker}: its goal is to gain an advantage over competitors by damaging company reputations.
\end{itemize}

\subsection{Man in the Middle attack} 
\label{sec:MitM}

\begin{figure}[htb]
    \centering
    \includegraphics[width=\columnwidth]{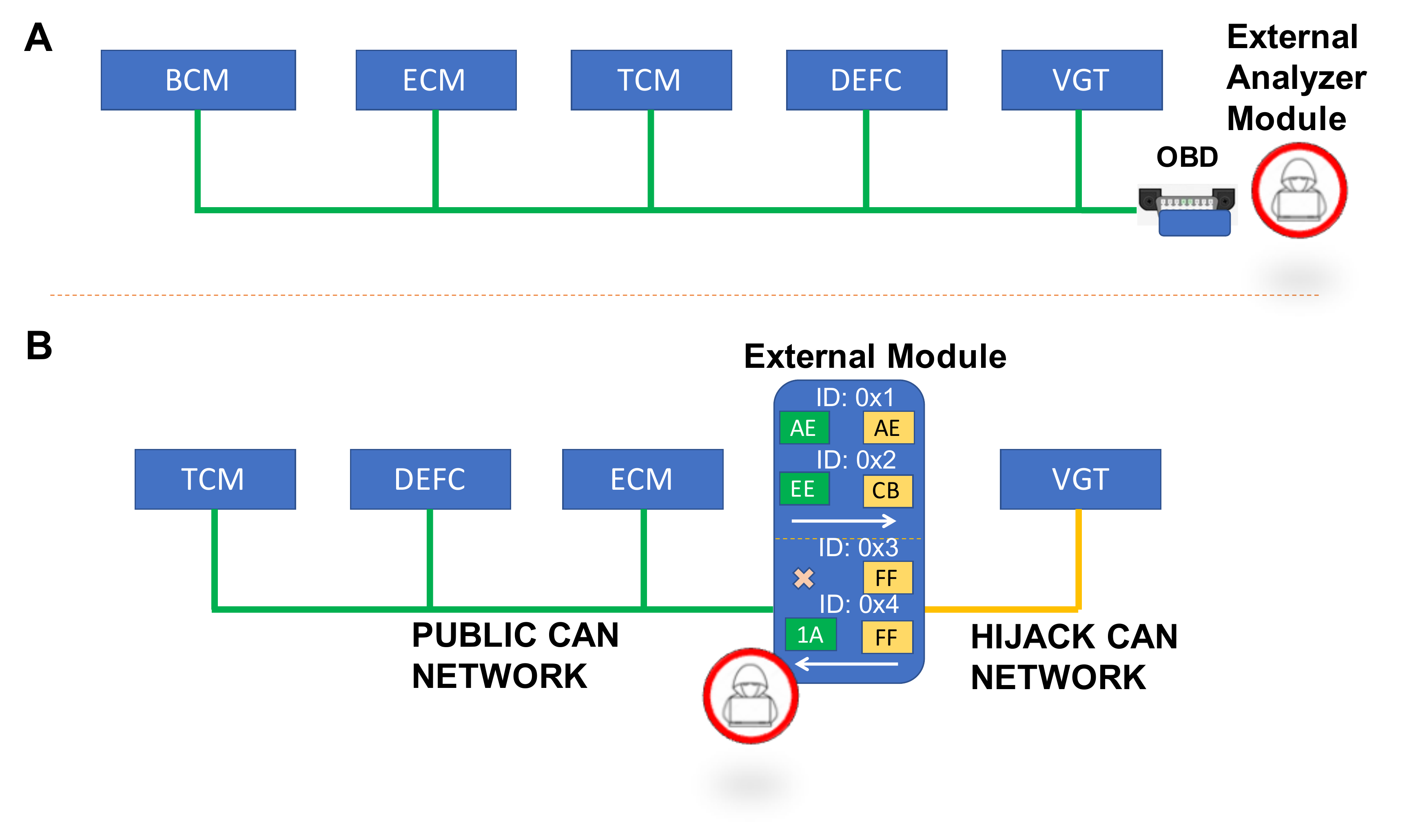}
    \caption{Man in the Middle attack schemes}
    \label{fig:MitM}
    \end{figure}

The Man in the Middle (MitM) attack is the preferred exploit implemented by vehicle owners. It is usually implemented resorting to external devices that create malicious CAN gateways. Figure~\ref{fig:MitM} shows the two standard settings for this attack: (A) exploiting the OBD port and (B) placing an external CAN module downstream to the victim module.

In the more straightforward implementation, the attacker connects an external analyzer to the OBD port to gain control over several diagnostic services (\autoref{fig:MitM}-A). The external analyzer can sniff the CAN network traffic, inject new CAN frames, and modify existing frames.

To gain additional power, the attacker must connect an external CAN gateway downstream to the victim ECU (\autoref{fig:MitM}-B). The malicious CAN gateway physically splits the network into two portions. Data frames generated by devices placed in the public CAN network can be conditioned before transiting to the hijack network where the victim ECU is located and vice versa. As an example, in \autoref{fig:MitM}-B, two frames sent to the Variable Geometry Turbine (VGT) are processed by the malicious gateway. Frame \texttt{ID 0x2} is corrupted while frame \texttt{ID 0x1} remains unaltered. Also, VGT generates two frames. The malicious gateway suppresses frame \texttt{ID 0x3} and forwards a modified version of the frame \texttt{ID 0x4}. Overall, with this configuration, an attacker can: 
\begin{itemize}
\item Intercept and then suppress specific messages;
\item Inject messages to emulate functionalities;
\item Intercept and then modify messages with corrupted data. 
\end{itemize}

Creation of new messages or modification of existing messages is possible with a \emph{direct attack} only when CMAC is not implemented or disabled. In all other cases, the attacker must execute an \emph{indirect attack}. In this case, the attacker performs a reply attack to bypass the CMAC signature by sniffing the network and reusing existing CAN messages \cite{Replyattack}. Reply attacks are easy to implement when a rolling counter is not applied to the exchanged messages. 

MitM attacks have a significant impact on warranty costs. Tampering with the vehicle parameters increases vehicle damage risks. In case of damage, the external devices used to mount the attack can be easily removed, making it impossible to prove a tampering action that would lead to a loss of warranty.

\subsection{Automotive Cyber-Security Key Provisioning Infrastructure}
\label{sec:key-exchange}

CMAC signatures guarantee the authenticity and integrity of CAN messages in automotive applications for all safety-critical and sensitive ECUs. The ECU security hardware architecture defines the number of keys required for CMAC calculation for each secure vehicle \cite{8762043}. CMAC calculation is a computation-intensive task that requires hardware acceleration. Therefore, the maximum number of secret keys a vehicle can handle strictly depends on the key length and the storage capability of the target Crypto Engine. The typical storage capability of a Crypto Engine today is around 256B. Assuming a 16B key size, it can potentially store 16 keys. It is expected that the next generation Crypto Engines will increase their storage up to 1 Kbyte, thus accommodating 64 16B keys. 

Carmakers must properly handle these secrets. Let us consider a big car-maker selling 10 Million secure vehicles per year.  If each car of the entire fleet uses a unique set of sixty-four 16B MAC  secret keys, the total amount of storage required to handle the keys would be approximately 9GB. This value may increase by a 3x factor by considering complementary information, such as Vehicle Identification Number (VIN) or module part-numbers.

\begin{figure}[ht!]
\centering
\includegraphics[width=\columnwidth]{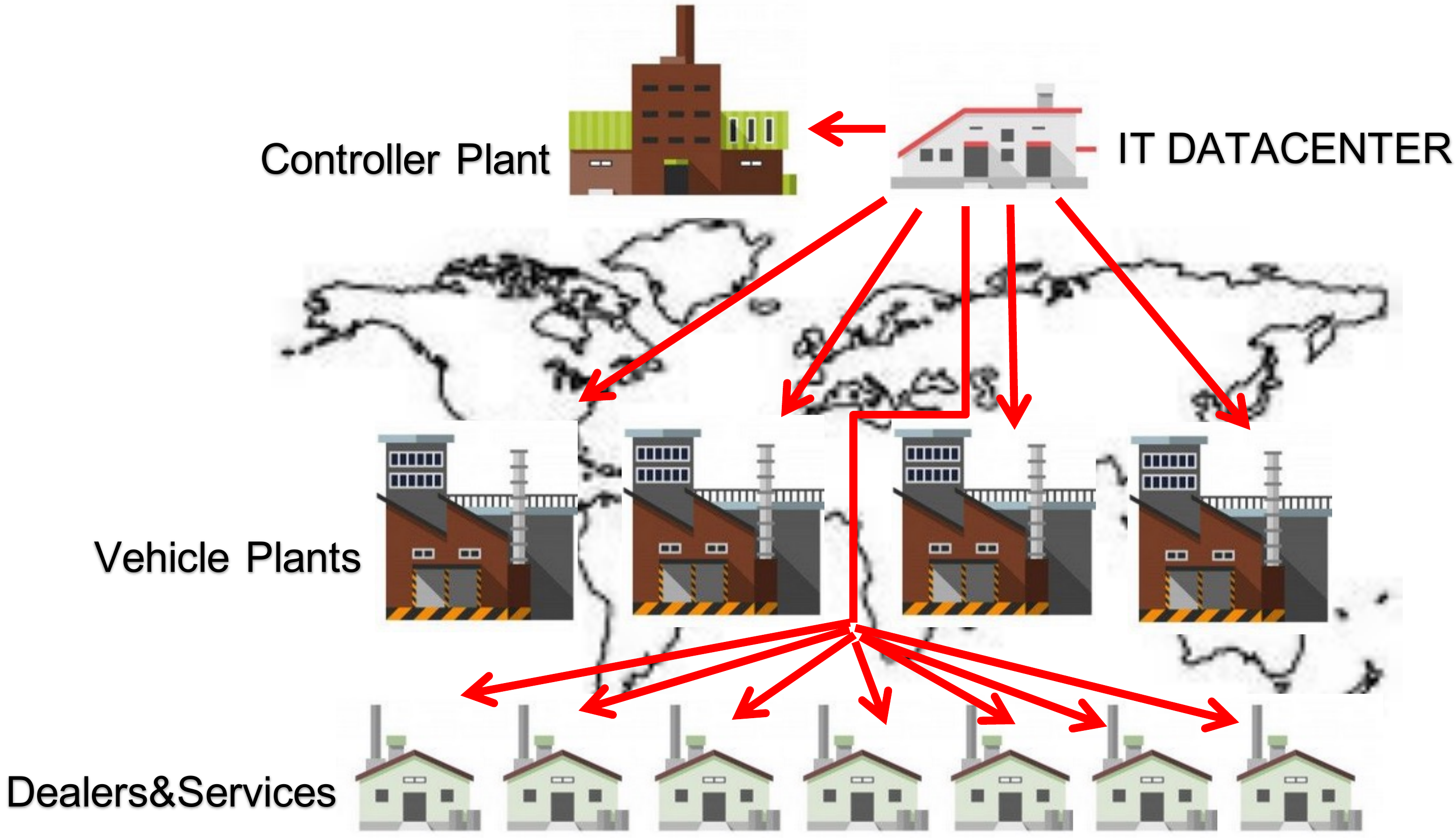}
\caption{Generic shared secret key proliferation}
\label{fig:9}
\end{figure}

Interestingly, these numbers do not represent a technical issue for an IT infrastructure. Nevertheless, key management requires significant security investments since data must be shared among different worldwide actors, including manufacturing plants, suppliers, services, and dealers (\autoref{fig:9}). It is not always easy to maintain trusted environments and avoid leakages in this context. Any violation compromises the entire vehicle fleet. Carmakers desire to dismiss the IT infrastructure having local key provisioning directly at the vehicle level, with a self-build method to mitigate the above risks.

\subsection{Denial of Service attacks} 
\label{sec:DoS}

The Denial of Service (DoS) attack is the preferred exploit for attackers that want to destroy a company's reputation. The attacker tries to gain public access to the CAN network to force a bus off or create a task overrun event. This is usually implemented by looking for infotainment system exploits leveraging the available apps or exploiting the presence of OBD Bluetooth devices associated with unofficial apps.

If a task overrun generates a CAN communication failure or a real-time deadline miss in automotive applications, the car's safety is compromised. Therefore, the vehicle must apply a safety recovery action with a potential impact on customers. For this reason, in a secure and safe architecture, a CAN gateway/firewall is usually inserted between the OBD port and infotainment system to the rest of the public CAN network. However, architectures that maximize the throughput of the CAN network increase the complexity of mounting DoS attacks.

\subsection{Hardware replacement attack}
\label{sec:introattack-model}

Vehicles embed several ECUs communicating through a CAN network. Software executed on these ECUs is a potential exploit for an attacker. A well-designed secure boot is among the most efficient protection against malicious software corruption in real-time ECUs \cite{8474730}. At each bootstrap, the system validates the signature of each memory segment. Moreover, code updates require an authentication mechanism to avoid the injection of potentially counterfeit software. 

In this scenario, upcoming market trends might favor attackers in their aim to neutralize boot signatures, granting unauthorized software to run in the system with potential hazards to the safety of the entire vehicle.

\begin{figure}[!ht]
\centering
\includegraphics[width=\columnwidth]{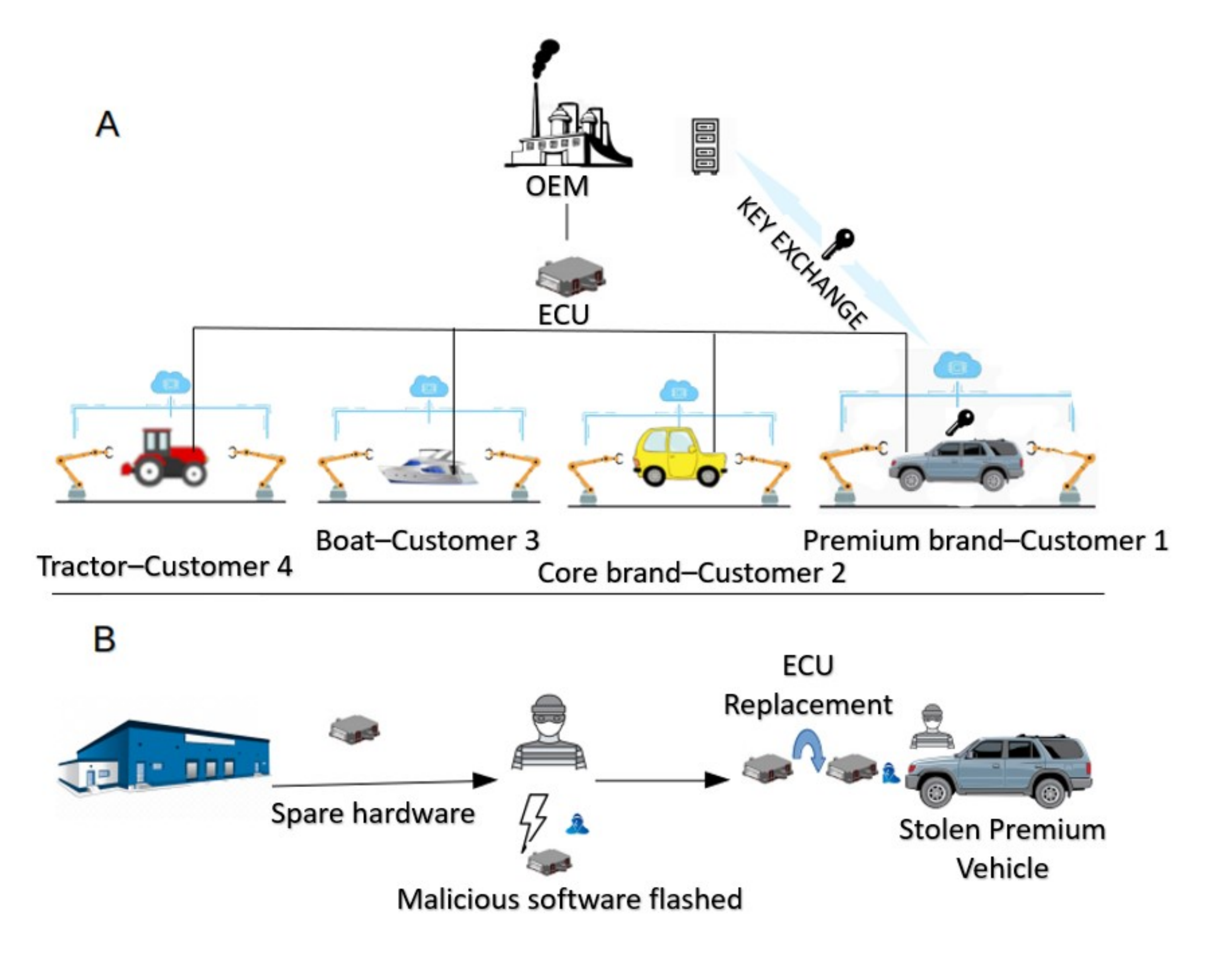}
\caption{Hardware replacement attack model: (a) the same ECU is exploited in several application domains and (b) an ECU can be easily reworked from one domain to another.}
\label{fig:overview}
\end{figure}

The automotive market pushes competition in terms of costs by exploiting the economy of scale. As depicted in \autoref{fig:overview}-A, an Original Equipment Manufacturer (OEM) delivers the same hardware platform to several customers who act in different heterogeneous domains with varying cyber-security requirements (e.g., automotive, marine, agriculture, general-purpose equipment) \cite{6542519}. A skilled attacker can easily rework unsecure ECUs sold in a market domain to serve another environment that adopts the same hardware platform (\autoref{fig:overview}-B). If this hardware replacement attack targets ECUs that require cyber-security in the target domain, code signature can be bypassed, thus allowing the execution of untrusted software. Therefore, hardware platforms must guarantee authenticity.

Physical Unclonable Functions (PUF) and Logic Locking, which are techniques proposed in the literature for hardware fingerprinting, are hard to be employed in the automotive domain. 

In vehicles, ECUs operate in an extensive range of environmental conditions (i.e., temperature, pressure, humidity) that may severely impact the PUF challenge's success \cite{6379926},\cite{9291788}. Moreover, external hardware or other control modules need to validate the challenges, hence defining an additional infrastructure similar to what is shown into \autoref{fig:9}. That increases the complexity of managing the authorized hardware part replacement events at services. While PUFs are very powerful in identifying every hardware device, the automotive domain is more interested in tracking control modules associated with a selected customer, application, or brand. 

Logical locking is a hardware technique based on integrating a locking mechanism into the circuit such that it produces faulty outputs whenever an incorrect key is provided \cite{7362173},\cite{9070188},\cite{8192439}. Logical locking is not well suited for automotive applications since the faulty outputs may generate hazards and violate the vehicle safety rules.

\section{Extended TAURUM P2T}
\label{sec:taurum-P2T}

\begin{figure}[ht!]
    \centering
    \includegraphics[width=0.8\columnwidth]{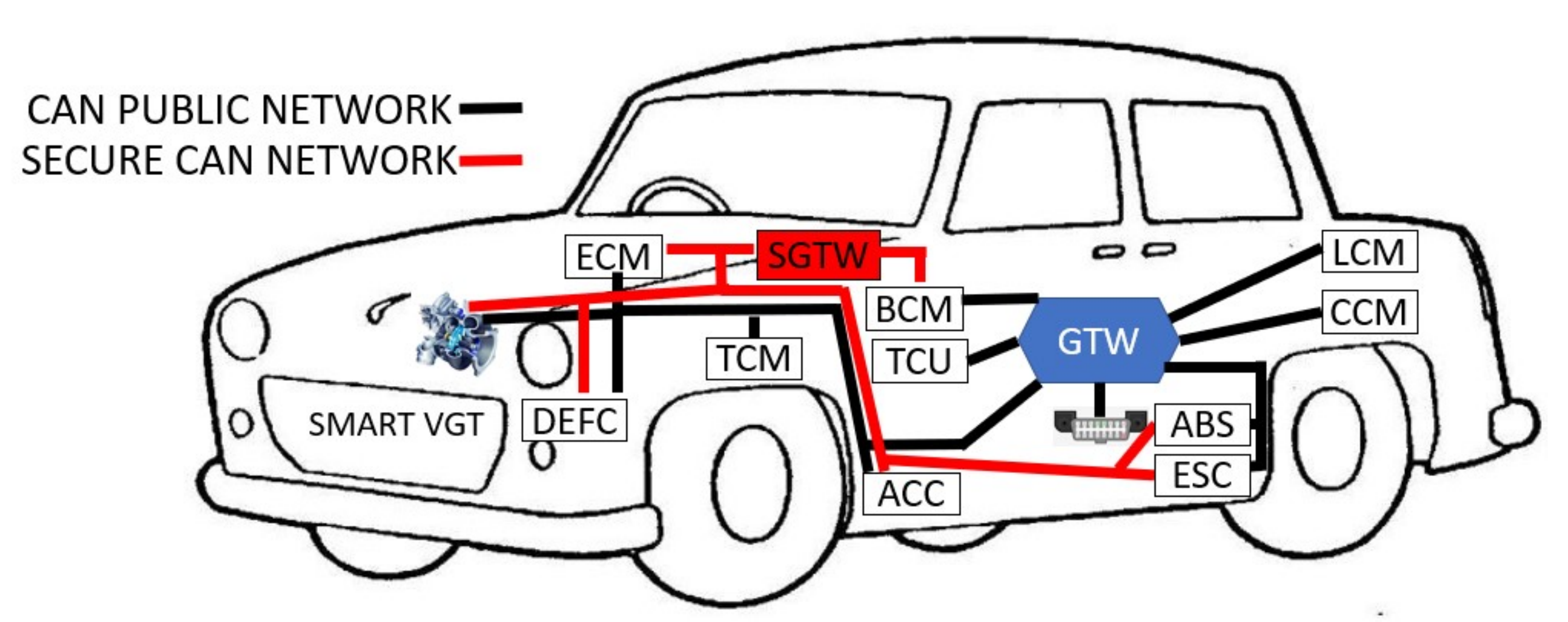}
    
    \caption{TAURUM P2T Advanced Secure CAN Network for Automotive.}
    \label{fig:6}
\end{figure}

EXT-TAURUM P2T is a secure infrastructure aiming at addressing the issues discussed in \autoref{sec:CAN-vulnerabilities}. EXT-TAURUM P2T is based on two independent CAN networks (\autoref{fig:6}). The \emph{Public CAN} network (depicted in black) transports the standard vehicle CAN traffic and is accessible through the standard CAN Gateway (CGTW). The \emph{Secure CAN} network (depicted in red) exchanges sensible information to handle shared keys, security violations, and signatures. Frames exchanged over the \emph{Secure CAN} network are encrypted, and the EXT-TAURUM P2T Secure Gateway (SGTW) guarantees controlled access to this network. It establishes privilege levels and manages secret keys required to compute MAC signatures. The main features provided by EXT-TAURUM P2T are:
\begin{itemize}
    \item a sharing key mechanism able to define isolated \emph{trust zones};
    \item a \emph{sub-domain} management of the bus for ensuring segregation;
    \item a \emph{rolling MAC secret key} infrastructure to implement a countermeasure to MitM and reply attacks;
	\item a \emph{dynamic key length adjustment mechanisms} and \emph{speculative MAC} calculation to maximize throughput and increase the complexity of DoS attacks;
	\item a \emph{challenge-response hardware authentication mechanism} to implement a countermeasure against hardware replacement attacks.
\end{itemize}

To be ready for the automotive industry, EXT-TAURUM P2T is entirely built, resorting to state-of-the-art cryptography and security standards.

\begin{figure}[htb]
    \centering
    \begin{subfigure}[t]{0.23\textwidth}
         \centering
             \includegraphics[width=\textwidth]{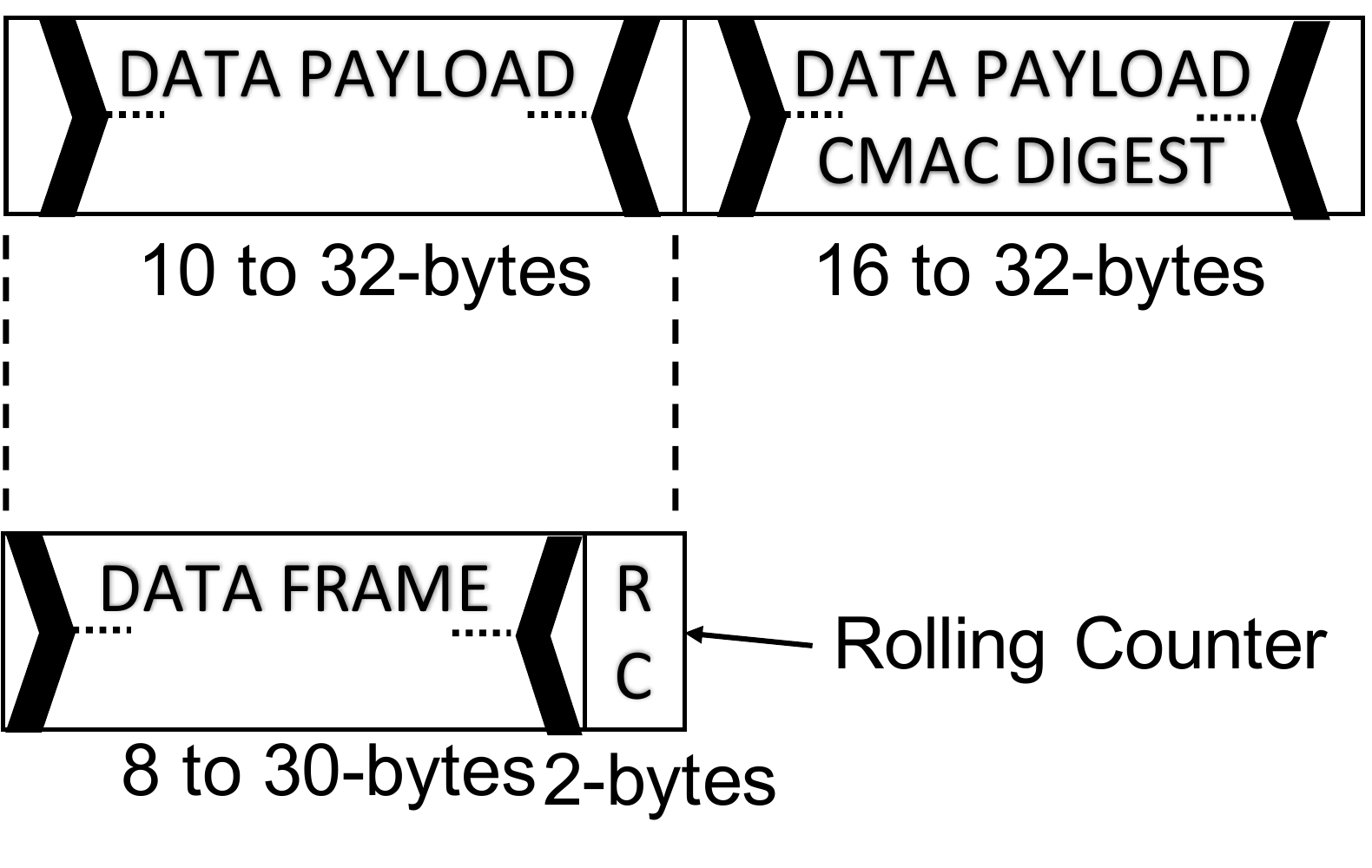}
         \caption{\scriptsize{PUBLIC CANF-FD Frame: a payload 26 bytes to 64 byte long divided into  data payload and CMAC digest required for authenticity and integrity purposes}}
         \label{fig:canfd_a}
     \end{subfigure}
     \hfill
     \begin{subfigure}[t]{0.23\textwidth}
         \centering
         \includegraphics[width=\textwidth]{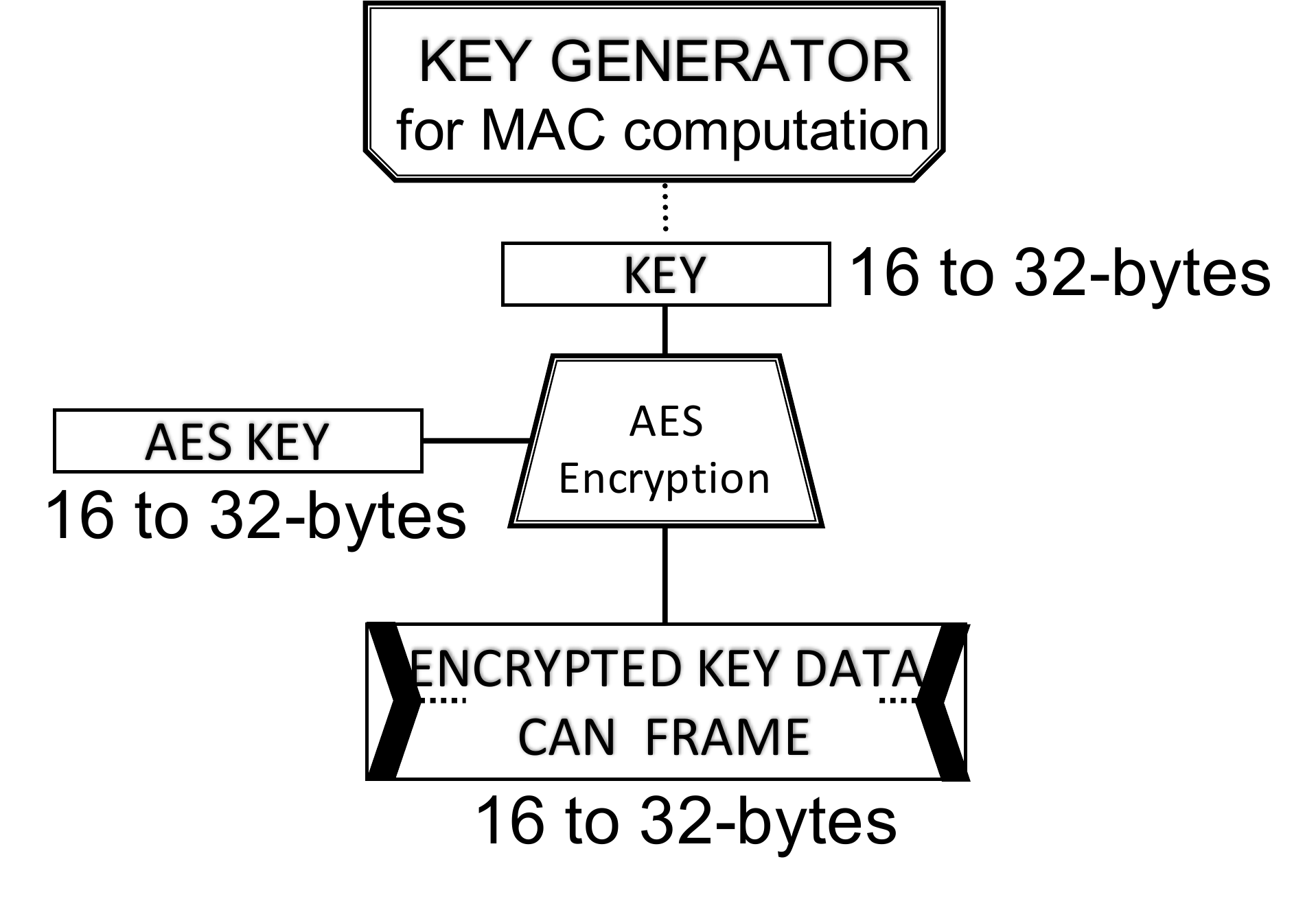}
         \caption{\scriptsize{SECURE CAN-FD Frame: encrypted message sent by the SGTW to all the secure modules for key update and security management purposes.}}
         \label{fig:canfd_b}
     \end{subfigure}
    \caption{TAURUM P2T Frames \label{fig:15}}
\end{figure}

EXT-TAURUM P2T requires a data rate of up to 8 Mbps and 64B data frames to be implemented. These requirements are met by the CAN-FD extension of the original CAN bus protocol \cite{6884472}.
The two communication networks transport the two classes of data frames depicted in \autoref{fig:15}: the Public CANF-FD frame transmitted over the Public CAN (\autoref{fig:canfd_a}) and the Secure CAN-FD frame transmitted over the Secure CAN (\autoref{fig:canfd_b}).

The integrity and authenticity of Public CANF-FD frames are guaranteed by including a CMAC digest of the transmitted data payload in the frame. CMAC signature computation is a time-consuming task.

Profiling CMAC computation time using real automotive hardware (see Section \ref{sec:results}) highlighted that the most secure architecture able to respect worst-case throughput constraints could employ 256 bit for data and 256 bit as CMAC digest. This configuration is the most protected from a cryptography standpoint, requiring secret key updates at a slow rate. Similar security levels can be obtained with fewer digest bits at the price of an increased key update rate, thus allowing to trade-off between digest's length and key updates. CMAC might not be enough to protect data transmitted over the public CAN. Messages containing steady-state information remain unchanged over time, favoring the implementation of replay attacks. For this reason, Public CANF-FD frames reserve two bytes for implementing a rolling counter protecting the system from these attacks \cite{rollingcounter}.

\begin{figure}[ht!]
\centering
\includegraphics[width=0.8\columnwidth]{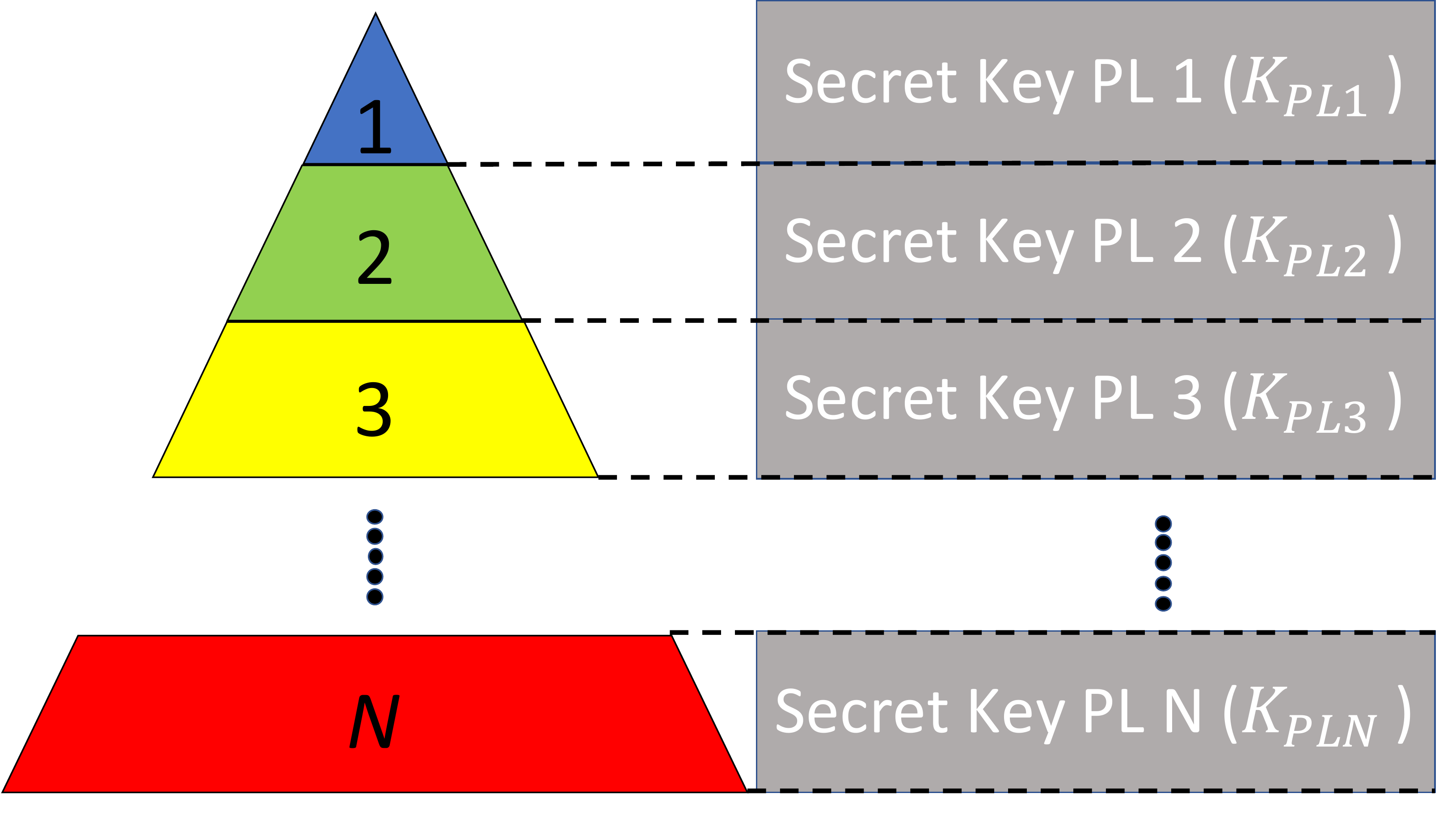}
\caption{TAURUM P2T Privilege Hierarchy Block Scheme. Lower numbers indicate higher privilege levels. Level 1 usually represents the SGTW.}
\label{fig:privileges}
\end{figure}

CMAC digest computation requires sharing a secret key between the sender and the receiver ECU. In a complex vehicle infrastructure, secure communication requirements are not orthogonal among all ECUs. Every ECU requires securely communicating to local groups of other ECUs depending on their executed tasks. Communications between tasks running on different ECUs must be segregated whenever possible to increase security. To handle this scenario, EXT-TAURUM P2T introduces the concept of privilege levels (PL) in the communication (\autoref{fig:privileges}). 

Privilege separation is a fundamental security feature introduced by EXT-TAURUM P2T. Every secure ECU also called a secure node (SN), is associated with a PL. Each PL holds a dedicated secret key ($K_{PLi}$) used for MAC signature computation between tasks executed at the same level. Privileges are organized in a hierarchy with low numbers indicating higher privileges. An ECU working at $\mathrm{PL_i}$ holds all secret keys form $\mathrm{PL_i}$ to $\mathrm{PL_N}$ (i.e., $K_{PLi}, K_{PLi+1}, \ldots , K_{PLN}$). It, therefore, can communicate with its counterparts at the same PL or with counterparts at lower PLs. 

With this mechanism, EXT-TAURUM P2T implements security segregation. Suppose an attack on an ECU succeeds in compromising its secret keys. In that case, only the ECU privilege level and all lower levels will be compromised until the activation of recovery countermeasures or update of the private keys takes place. Communication at higher PLs remains active, thus minimizing the attack's impact on the vehicle's functionalities.

EXT-TAURUM P2T privilege separation also implements an additional feature to handle specific vehicles security requirements. Road vehicles are often equipped with so-called secondary controllers. Usually, these modules have reduced hardware capability for meeting security requirements (e.g., key length restrictions). Directly connecting these devices to the entire network would decrease the overall security of the whole system. To avoid this, EXT-TAURUM P2T exploits PLs to define so-called security sub-domains. In a security sub-domain, the strength of the secret keys can be reduced (e.g., 8B or less) to better fit the system throughput constraints, helpfully allowing other parts of the system to work with more robust protections. This requires a more frequent update of the secret keys in sub-domains using shorter keys. 

\begin{figure}[ht!]
\centering
\includegraphics[width=\columnwidth]{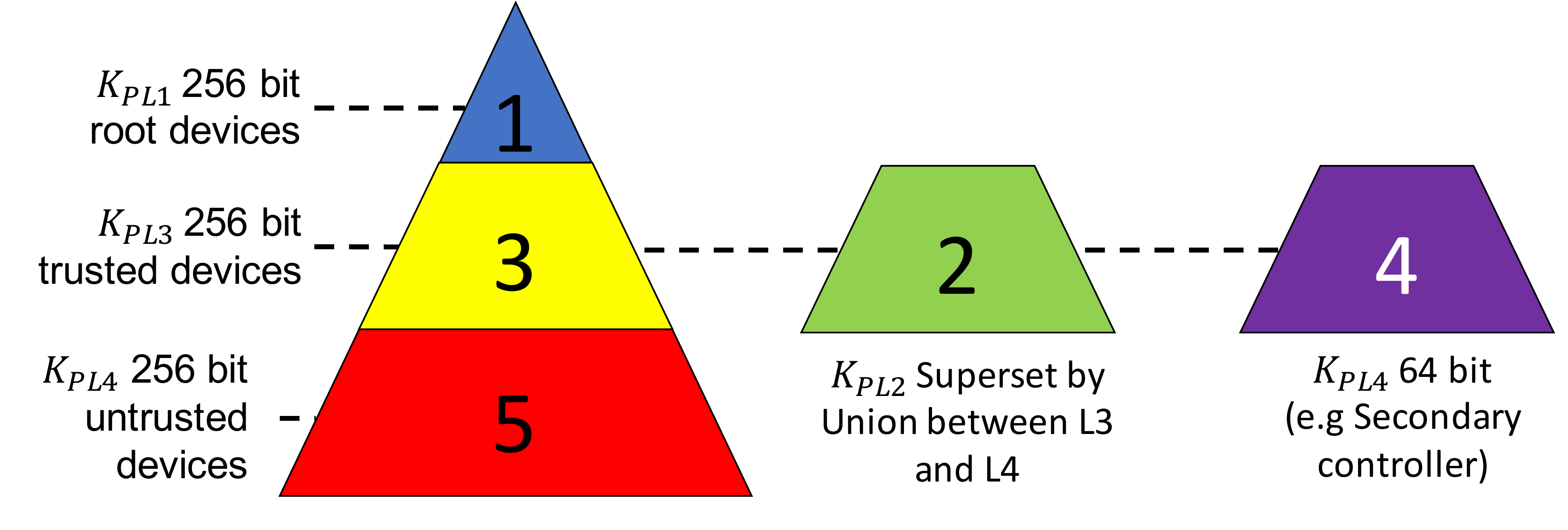}
\caption{TAURUM P2T Privilege Hierarchy with different key size. \label{fig:privilege-gateweay}}
\end{figure}

\autoref{fig:privilege-gateweay} provides an example of how sub-domains can be exploited. In this example, the SGTW works at level 1, while all non-critical devices of the network work at level 5. All safety-critical modules work at level 3, except a secondary controller with limited computing power that works at level 4. Finally, level 2 is associated with a sub-domain gateway module, thus keeping the secondary controller isolated from the other nodes of the CAN network. 

The way privilege levels are assigned is application-dependent and aims to fulfill the architecture's security requirements.


\subsection{Secure CAN and Key Provisioning }
\label{sec:keyprov}

The role of the EXT-TAURUM P2T secure CAN is to provide a secure channel to implement key provisioning and therefore share the secret keys ($K_{PLi}$) required by all SNs for CMAC digest calculation.

Communication on this channel must be fully secure and guarantee confidentially integrity and authenticity. State-of-the-art symmetric cryptography based on the Advanced Encryption Standard (AES), implemented with the Cipher Block Chaining (CBC) modality, represents the best approach to secure this communication channel \cite{AES}. The same PL secret keys ($K_{PLi}$) used for CMAC digest calculation are also used for encrypting communication at different PLs on the secure CAN network. To keep a high level of security, these secret keys are periodically rolled. The rolling time and the digest size are parametrized to ensure the highest flexibility.

This secure communication infrastructure setup requires a mechanism to distribute the secret keys to the different ECUs. As discussed in \autoref{sec:key-exchange}, the secret key distribution infrastructure is among the main challenges for carmakers in developing a large fleet of connected vehicles. EXT-TAURUM P2T removes this bottleneck by introducing a mechanism to generate all secrets on-board through the SGTW and securely share them with all connected nodes. This solution reduces the need to find trusted users and sustain a secure infrastructure.

\begin{figure*}[!ht]
    \centering
    \includegraphics[width=0.6\textwidth]{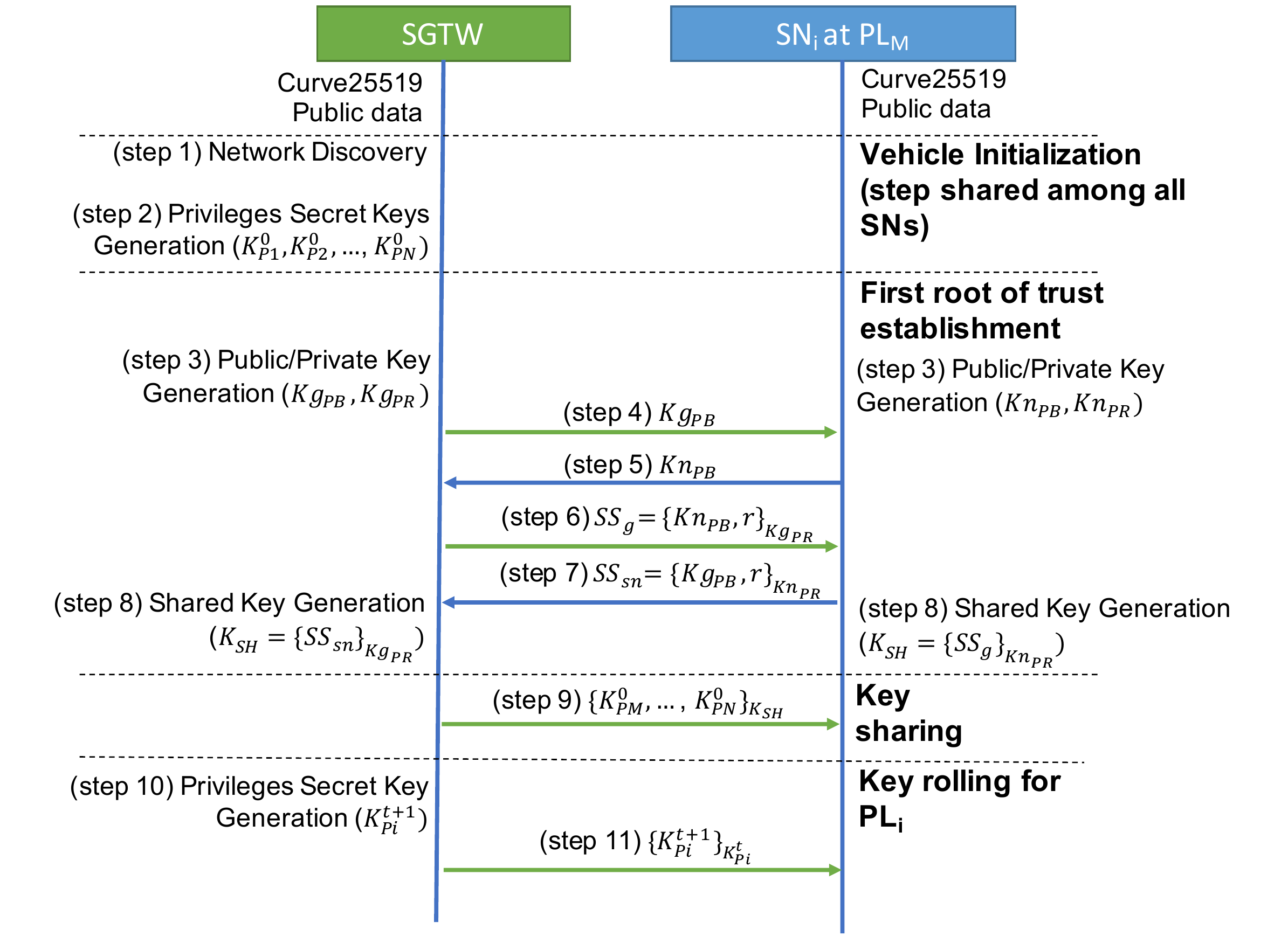}
    
    \caption{TAURUM P2T Secure CAN key provisioning protocol based on symmetric cryptography.}
    \label{fig:7}
    \end{figure*}

\autoref{fig:7} outlines the EXT-TAURUM P2T key provisioning protocol. During the first vehicle initialization at the plant (step 1), the SGTW performs a network discovery phase to map all SNs connected to the Secure CAN (i.e., those that require exchanging CMAC signed frames on the public CAN). It then generates using its local Crypto Engine the first set (time 0) of all PL secret keys ($K^{0}_{PL1}, \cdots, K^{0}_{PLN}$) and securely stores this information in its internal memory (step 2). After a complete network discovery, the SGTW handles the key provisioning node by node.

To establish the first root of trust between the SGTW and a given secure node ($\mathrm{SN_i}$) with privilege $\mathrm{PL_M}$, EXT-TAURUM P2T resorts to elliptic-curve cryptography (ECC) \cite{Koblitz04guideto}. Every ECU connected to EXT-TAURUM P2T stores the same curve as public data in its flash. Curve25519 has been selected since it is one of the fastest ECC curves enabling it to fit hard real-time constraints, it offers 128 bits of security (256 bits key size), and any known patents do not cover it \cite{10.1007/11745853_14}. 

ECC shared keys are used to provision MAC secret keys during the network's initialization or when an attack is detected. They make it possible to build a secure point-to-point network between the SGTW and each ECU. 

The SGTW and the SN start the establishment of the first root of trust (step 3) by generating a public/private key pair ($(Kg_{PB},Kg_{PR})$ for the SGTW and $(Kn_{PB},Kn_{PR})$ for the secure node). The SGTW uses a different key pair for every node. The SGTW and SN exchange their public key (steps 4 and 5) and use it to build two shared secrets ($SS_g$ and $SS_{sn}$), adding a nonce to the received public key. After encryption, these secrets are exchanged using the local private keys (steps 6 and 7). The shared secrets are used to generate the first shared key $K_{SH}$ (step 8). This shared key is used to securely transfer the secret keys starting from $\mathrm{PL_M}$ (the PL of the SN) down to $\mathrm{PL_N}$ ($K^{0}_{PLM}, \cdots, K^{0}_{PLN}$ in step 9). At this point, the node holds the secret keys and can start communicating with other nodes on the public network using CMAC signed frames. 

Generated keys are valid for a limited time frame. Each PL sets a rolling timer to decide when to roll its related key. Whenever the rolling key time of $\mathrm{PL_i}$ expires, the SGTW generates a new key (step 10) and then transmits the new key to all nodes connected to that level using the previous key. The secret key update is not only time-based but can also be event-based. An update can be forced by a specific event, like init, shutdown controller procedure, etc. 

\begin{figure}[!ht]
    \centering
    \includegraphics[width=\columnwidth]{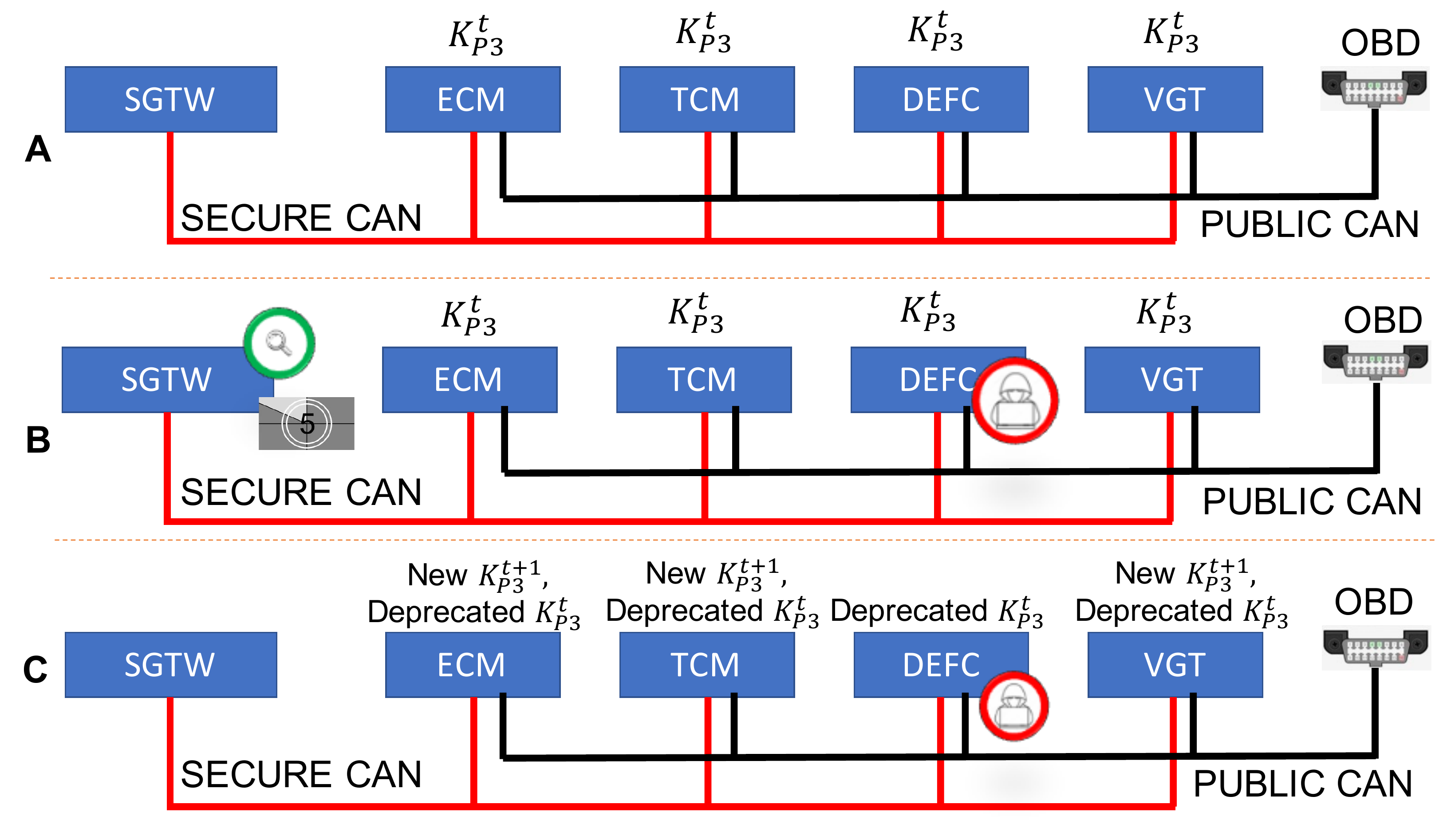}
    
    \caption{TAURUM P2T Secret Key Deprecate Status.}
    \label{fig:12}
    \end{figure}

EXT-TAURUM P2T implements a deprecated key functionality. When the violation of an ECU is detected, the SGTW can mark the related PL secret key as deprecated. \autoref{fig:12} shows an example of this mechanism. Starting from a valid condition with several ECUs connected at $\mathrm{PL_3}$ (\autoref{fig:12}A), the SGTW detects a compromised DEFC module (\autoref{fig:12}B). The secret key for $K^{t}_{SH}$ is then marked as deprecated (\autoref{fig:12}C). All ECUs connected at the same PL or higher are informed and receive a new key $K^{t+1}_{SH}$ encrypted using their $K_{SH}$. This isolates the compromised node on that level through privilege downgrading.
    
EXT-TAURUM P2T also includes a Short Secret Key mode providing each SN with an additional short key (e.g., 16B) in specific conditions. Forcing the network to work with shorter digests and keys saves throughput and computation resources. This mode helps to gain extra hardware resources for counterattacking or managing high throughput peaks.

\begin{figure}[!ht]
    \centering
    \includegraphics[width=8cm]{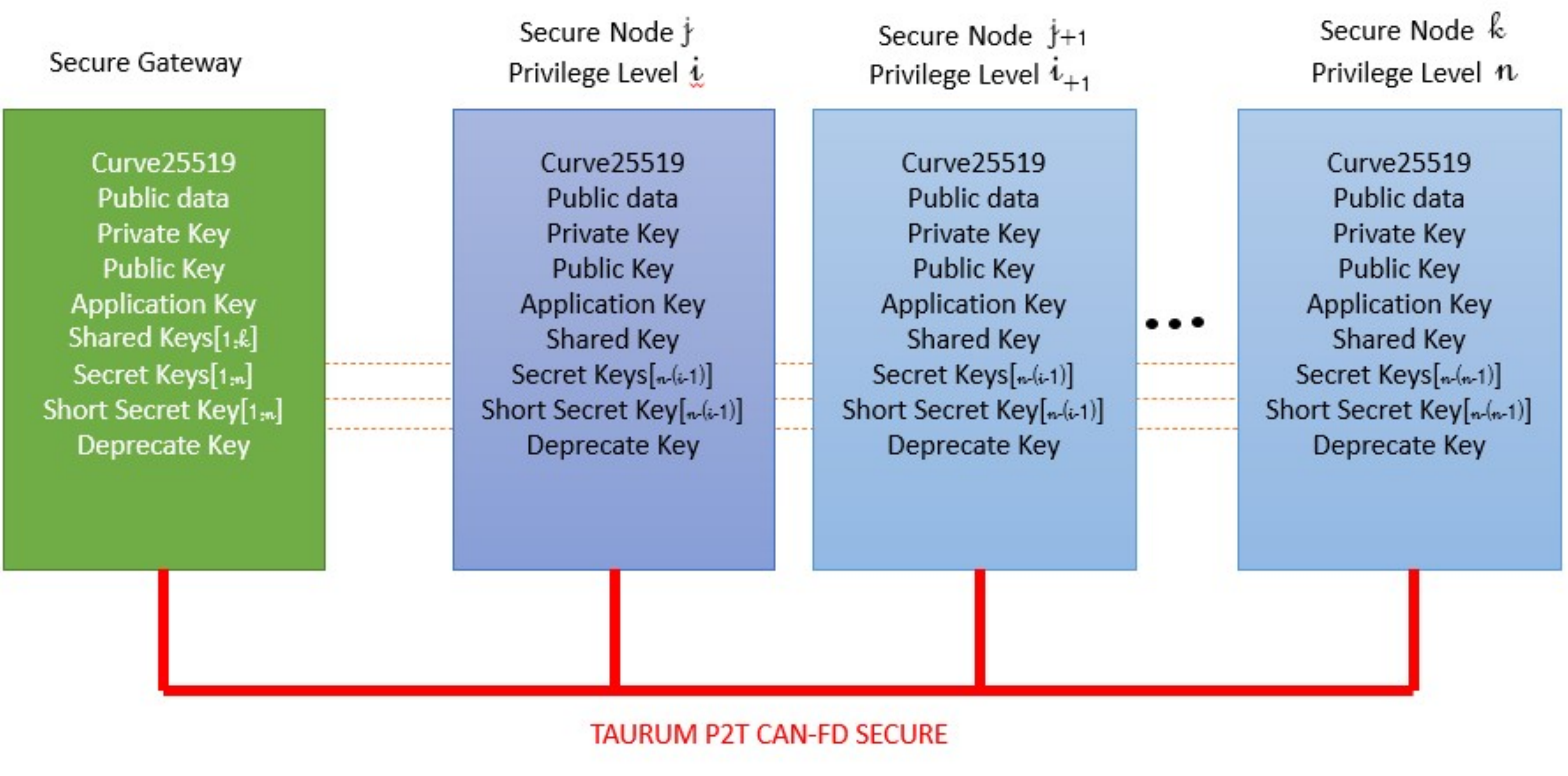}
    \caption{Summary of EXT-TAURUM P2T shared secrets}
    \label{fig:8}
    \end{figure}

To summarize, \autoref{fig:8} shows the secret keys that every module must handle in an EXT-TAURUM P2T architecture. EXT-TAURUM P2T centralizes hardware resources into the SGTW, allowing for a more flexible and lighter security resource into the rest of the connected modules. All controllers can implement a minimal encryption function with limited storage capacity.

\subsection{Speculative MAC calculation}
\label{sec:speMac}

The throughput is a sensitive parameter inside real-time systems, as discussed \autoref{sec:results}.  Security mechanisms, particularly CMAC computation, profoundly impact the system's throughput.  Attackers can exploit communications peaks to generate DoS attacks.  Therefore, reducing the network traffic in normal conditions is essential to have a margin when handling critical situations. To support this goal, EXT-TAURUM P2T introduces a speculative MAC computation mechanism to optimize the CPU load in connected ECUs, thus avoiding missing real-time deadlines during critical transient conditions.

\begin{figure}[!ht]
    \centering
    \includegraphics[width=9cm]{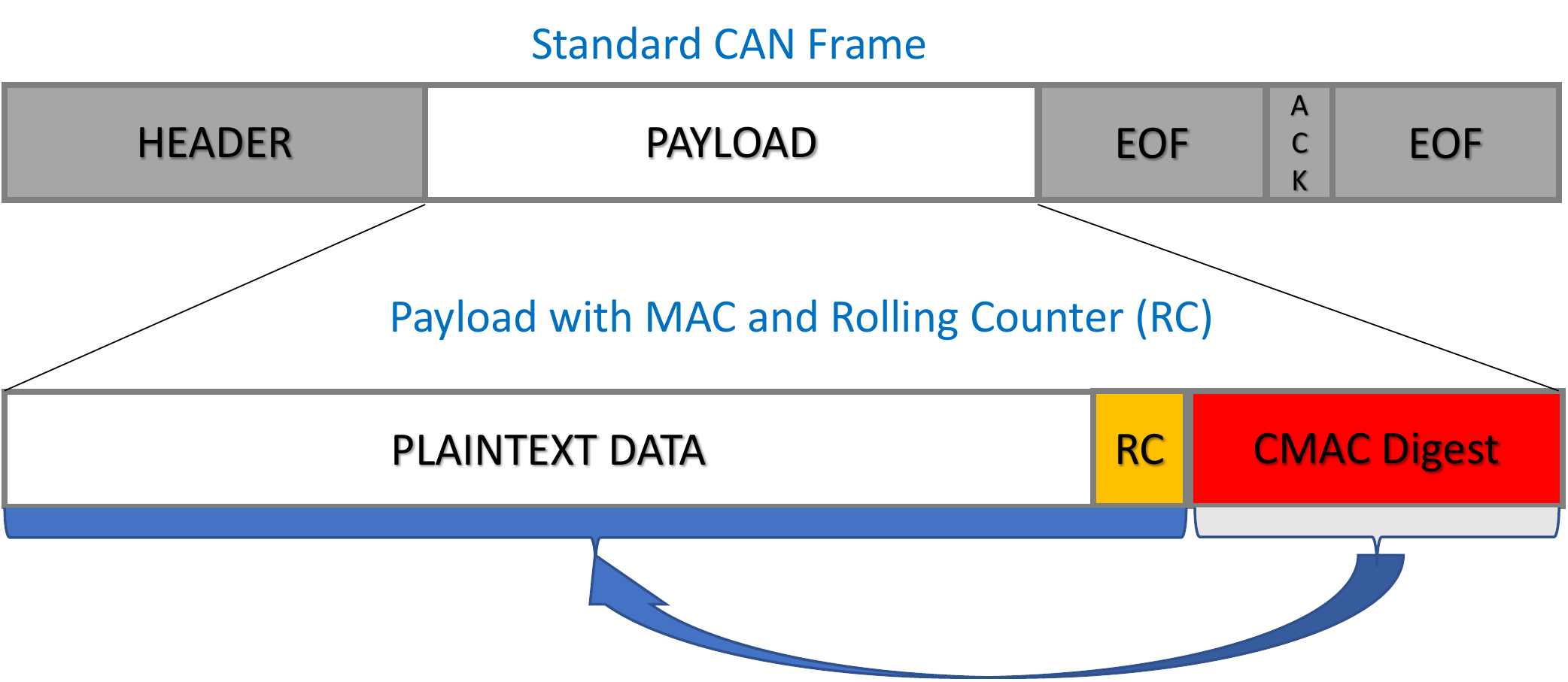}
    \caption{Use of MAC in CAN frames to guarantee integrity and authenticity}
    \label{fig:MACframe}
    \end{figure}

To understand how this mechanism works, let us start with a quick overview of how CMAC is used in CAN communication to guarantee the integrity and authenticity of a transmitted frame. To avoid reply attacks, the frame transmitter computes a signature (CMAC digest) of the plaintext data concatenated with a rolling counter. The plaintext data, the rolling counter, and the CMAC digest are embedded in the CAN frame payload and transmitted over the CAN network (\autoref{fig:MACframe}).  Before using data contained in a frame, the receivers must calculate the CMAC digest again and compare it with the one included in the transmitted frame.  If the two digests are the same, the integrity and authenticity of the CAN message are verified, and the frame can be used; otherwise, it is discarded and considered unauthorized.  The system moves in a recovery mode when a receiver often gets CAN frames with an invalid digest.  The system proceeds to a recovery mode in this second situation, depending on the function connected with the transmitted frame.

    \begin{figure*}[!ht]
    \centering
    \includegraphics[width=\textwidth]{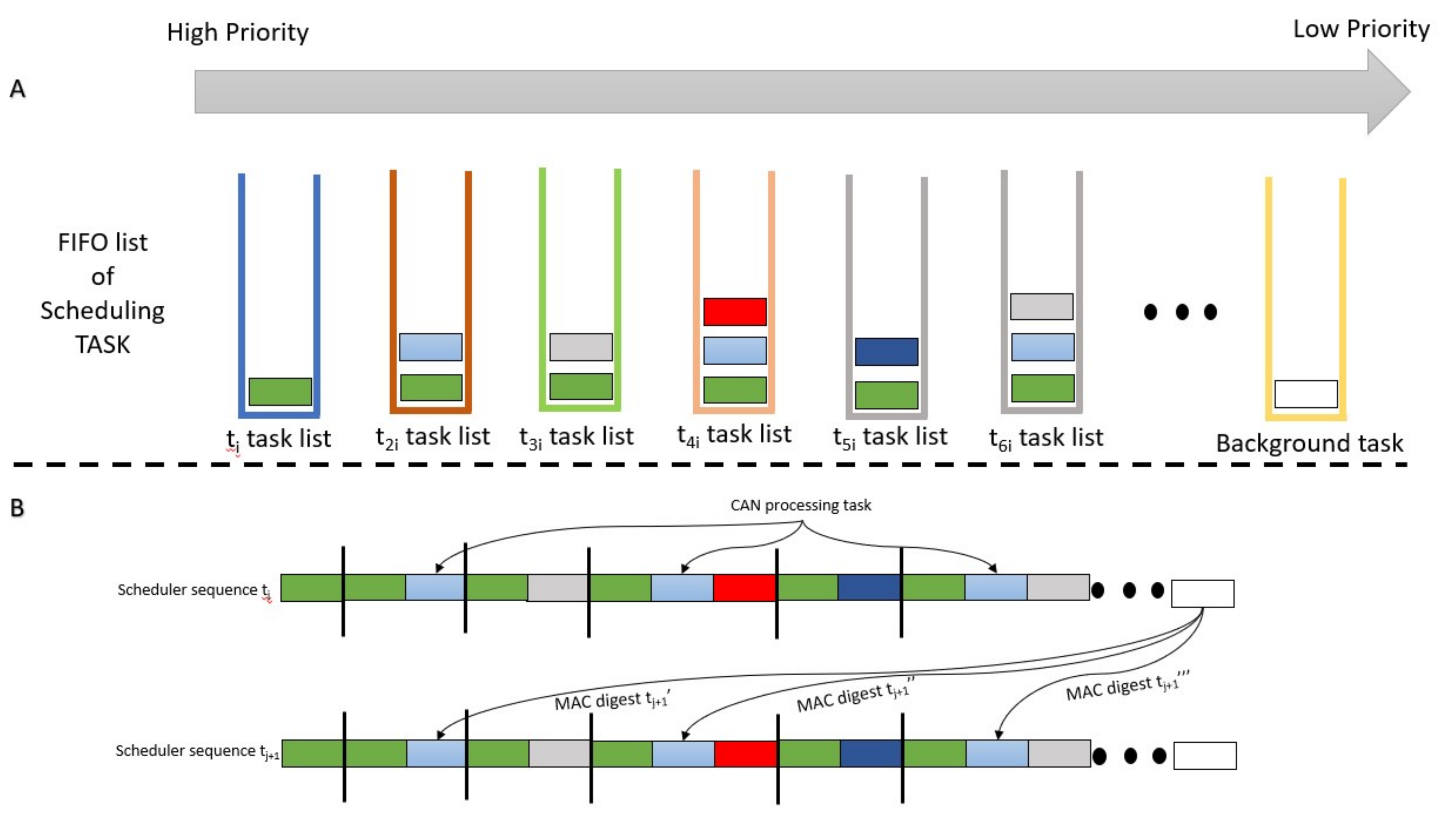}
    \caption{EXT-TAURUM P2T Speculative MAC calculation implemented in generic RTOS}
    \label{fig:oseksch}
    \end{figure*}

CAN data frames usually transport information obtained from the measurement of physical signals (e.g., temperature, pressure, rotation speed, etc.). While some of these measurements continuously change over time, other measures have slow changes and remain steady when considering short periods. Typically, steady-state signals are part of the following domains: temperature (e.g., Air Ambient temperature frame), atmosphere pressure  (e.g., ambient pressure frame),  voltages (e.g., battery voltage when generator's contribution is none), etc. 

In this context, it is possible to predict the future data frame payload and understand if there will be a difference or not,  just monitoring specific system parameters. In those cases, the rolling counter introduced to avoid reply attacks is the only change in consecutive CAN data frames. EXT-TAURUM P2T exploits this property to implement speculative MAC calculation thanks to the characteristics offered by the OSEK operating systems executed on automotive ECUs.

Figure \ref{fig:oseksch}-A represents a high-level view of how an OSEK operating system schedules tasks. Each colored rectangular is a task with its priority. Tasks at the same priority are scheduled according to a FIFO policy. When no task requires the CPU, the background task is executed. Figure \ref{fig:oseksch}-B shows the same task scheduling from a different perspective. Let us focus on tasks receiving and processing CAN data frames (light blue rectangles). Before using the information contained in a frame, these tasks must compute the CMAC digest and compare it with the one stored in the frame to perform message authentication. The speculative approach of EXT-TAURUM P2T delegates the digest computation for all frames containing steady-state measures to a low priority task (background task), keeping just the comparison instruction between the two digests in the original task. In \autoref{fig:oseksch}-B, at time $t_{j}$ the background task computes speculative MAC digests for steady-state frames that are used for message authentication at time  $t_{j+1}$. 

\begin{figure}[!ht]
    \centering
    \includegraphics[width=\columnwidth]{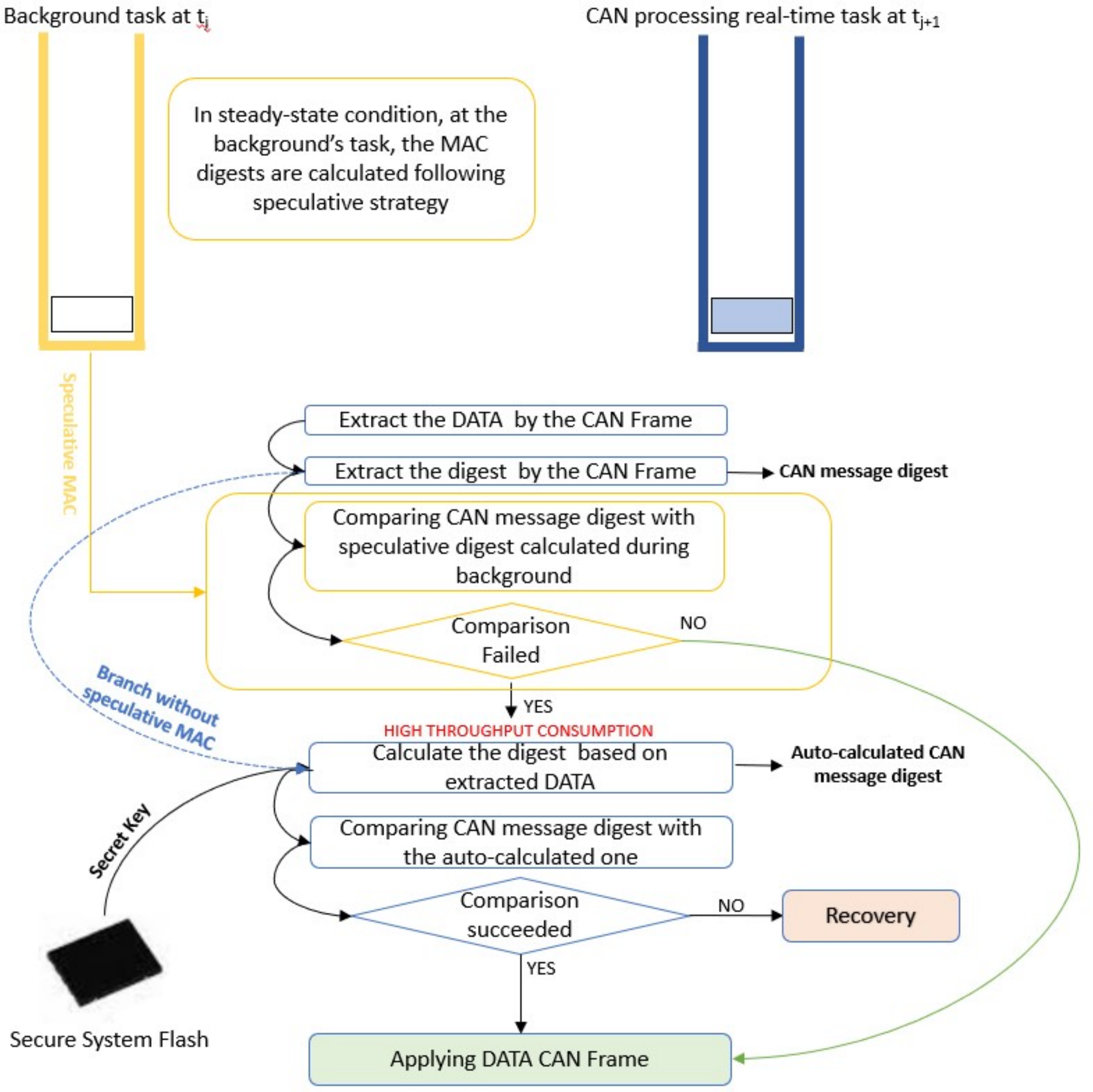}
    \caption{MAC Speculative Strategy block scheme}
    \label{fig:specMACflow}
    \end{figure}

It is essential to highlight that the introduction of the speculative MAC does not introduce any security threat to the system. The speculative MAC computation operates at the receiver's side following a flow summarized in Figure \ref{fig:specMACflow}. CMAC digests for CAN frames that likely transmit steady-state information (steady-state frames) are computed in a background task exploiting idle CPU time and stored for later use. When a frame arrives, the receiver first compares the frame digest with the speculative digest that is already available. If the comparison succeeds, it means the frame contains steady-state information, and the speculative MAC mechanism could predict it in advance. The frame can be considered secure and used for further computations. If this check fails, either the frame is corrupted, or the contained information is not steady-state, and therefore the speculative MAC was unable to perform a correct prediction. In this case, the receiver switches back to a standard validation flow. It extracts the plaintext and rolling counter from the frame and computes the MAC digests. It then compares it with the one stored in the frame to assess its integrity and authenticity. If this comparison succeeds, the frame can be used. Otherwise, it must be discarded.

Moving MAC computation to a background task has an enormous advantage. Its operations are not under real-time constraints and do not contribute to CPU real-time utilization. Furthermore, this approach allows also to solve safety's constraints, being safety put in a strong relationship with the real-time, and all tasks outside the real-time domain are considered without any impact on safety.

Speculative MAC computation is a valuable technique to mitigate secure hardware overload peaks.


\subsection{Hardware signature for branding system}
\label{sec:hwsignature}

EXT-TAURUM P2T provides a secure communication infrastructure to implement the hardware signature mechanism conceptually introduced in \cite{9525579}, able to avoid the hardware replacement attack described in section~\ref{sec:introattack-model}. 

To generate a compatibility discontinuity among hardware platforms integrated into different market subdomains, every carmaker buying parts from OEM must securely store a shared secret $\mathrm{K_{apk}}$ into every ECU (including the EXT-TAURUM P2T SGTW). This secret is unique for every carmaker and is used to verify the origin of the ECU.

\begin{figure}[!ht]
    \centering
    \includegraphics[width=\columnwidth]{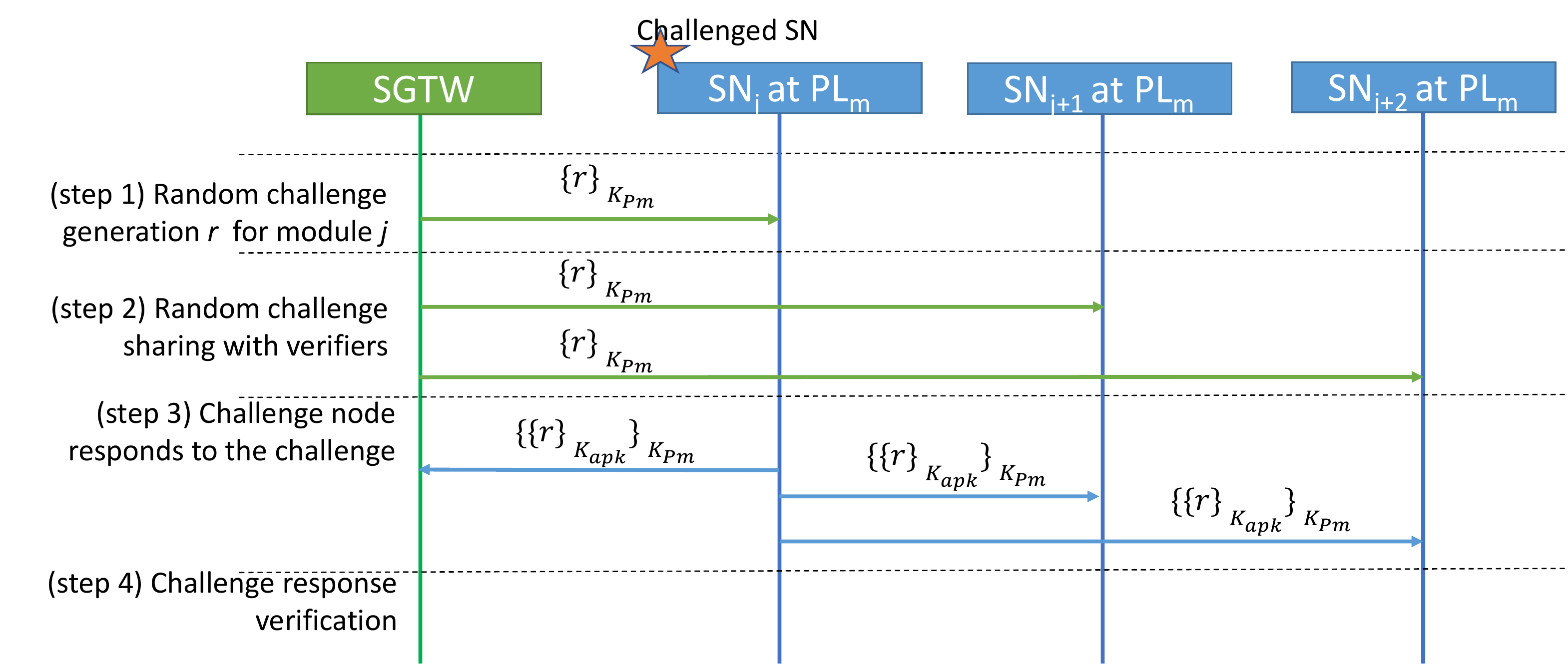}
    \caption{EXT-TAURUM P2T Hardware Signature with challenge response authentication}
    \label{fig:challenge_response}
    \end{figure}

During operation, the EXT-TAURUM P2T SGTW uses the Secure CAN network to periodically initiate a distributed hardware verification protocol depicted in \autoref{fig:challenge_response}. The verification process is local to every PL. Considering privilege level m, SGTW selects a target ECU to be verified randomly. It generates a nonce $r$ and sends it over the Secure CAN network encrypted with the corresponding PL key $\mathrm{K_{PLm}}$ (step 1). It then shares the same random challenge over the Secure CAN  network with all other ECUs working at the same PL (step 2).

At this stage, the challenged ECU must answer the challenge by encrypting $r$ using the carmaker secret key $\mathrm{K_{apk}}$ and sending this information back to the SGTW and all other ECUs at the same PL encrypted with the PL secret key $\mathrm{K_{PLm}}$. The SGTW and all ECUs that receive the challenge-response can act as verifiers, checking the correctness of the response. Suppose at least one ECU detects a violation. In that case, the EXT-TAURUM P2T key deprecation feature can be activated to isolate the complete PL subdomain in the network and initiate a recovery action to exclude non-authentic hardware to keep the system safe for a certain period before permanently invalidating the compromised module. In case of failed response without recovery mode triggered by SGTW means that SGTW is compromised too. Another way to identify a discredited SGTW is to monitor the challenged module selection. If nodes detect that a particular node has never been challenged is a symptom of a tampered network. 

The carmaker's secret key becomes the critical security actor in this condition. However, information leakage for a specific company does not become a threat for all companies with the same hardware platform since they all have a proper programmed secret key.

\section{Experimental results}
\label{sec:results}
\begin{figure}[!ht]
\centering
\includegraphics[width=0.7\columnwidth]{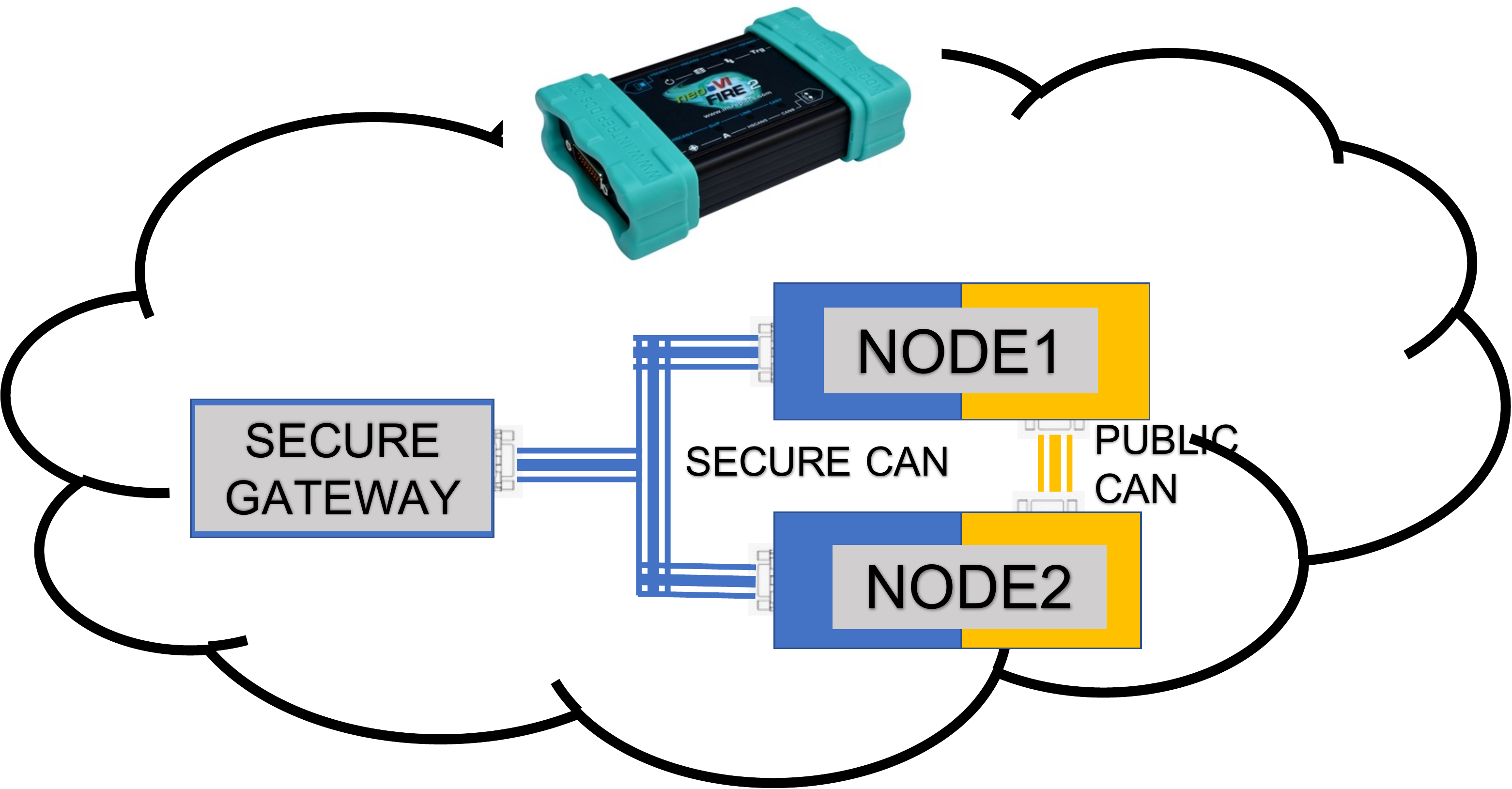}
\caption{EXT-TAURUM P2T simulation environment set up}
\label{fig:neovi2}
\end{figure}

The EXT-TAURUM P2T Secure CAN network concept was verified by simulating an authentic vehicle architecture, including the SGTW connected to two nodes. The implementation was based on the neoVI FIRE 2 Multi-Protocol Vehicle Interface produced by Intrepidcs \cite{NEOVI}. The device was configured with a CANF-FD baud rate of 500Kbit/s, and EXT-TAURUM P2T was configured to manage up to five PLs (\autoref{fig:neovi2}). The entire communication stack was built using Python.

The proposed simulation environment was used to experimentally validate the EXT-TAURUM P2T architecture, providing interesting information about the feasibility and performance of all security mechanisms. 

\subsection{Performance evaluation}

\begin{figure}[htb]
    \centering
    \begin{subfigure}[t]{0.23\textwidth}
         \centering
             \includegraphics[width=\textwidth]{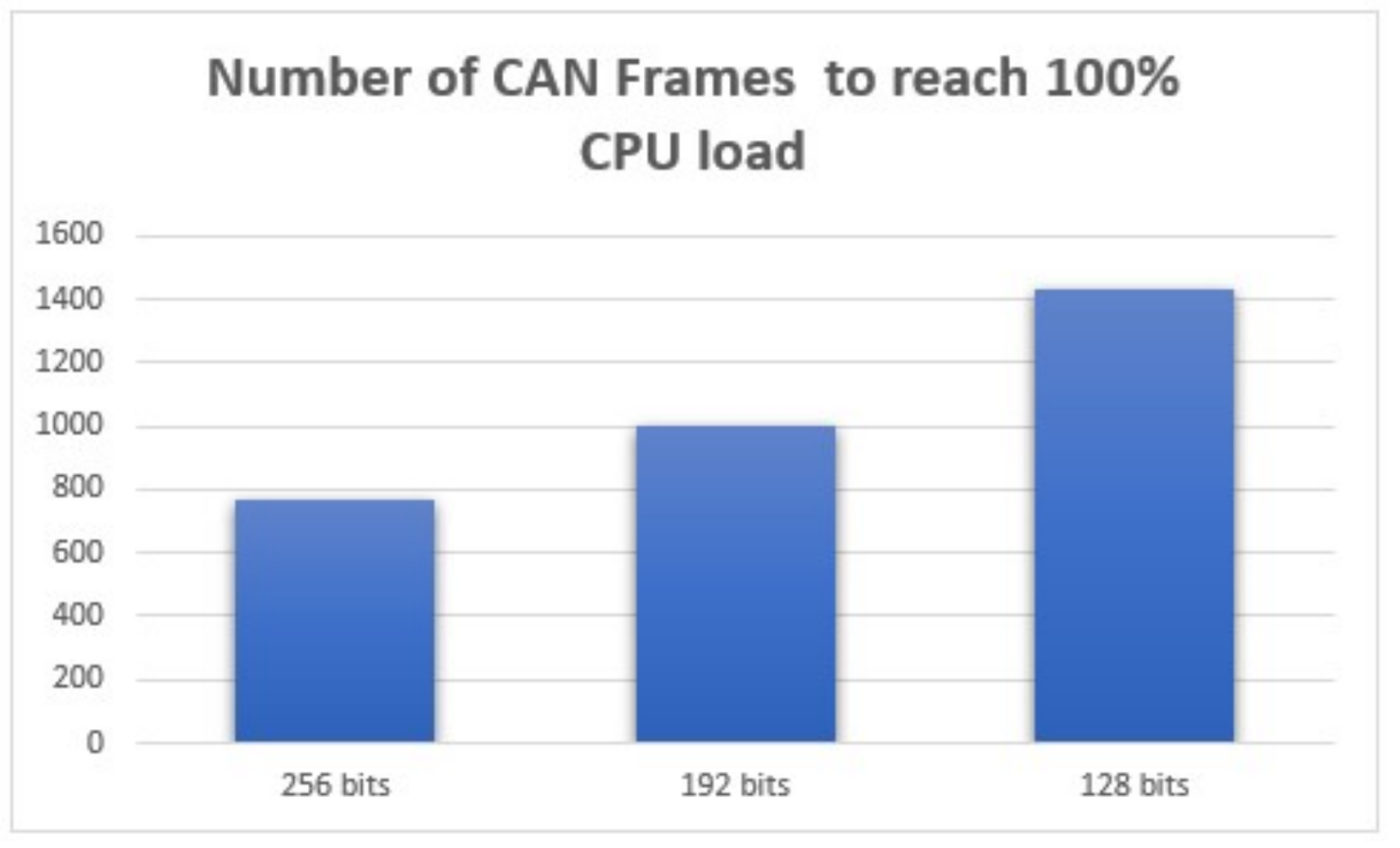}
         \caption{\scriptsize{Maximum number of frames to be processed to reach 100\% CPU utilization for MAC processing.}}
         \label{fig:maxnframes}
     \end{subfigure}
     \hfill
     \begin{subfigure}[t]{0.23\textwidth}
         \centering
         \includegraphics[width=\textwidth]{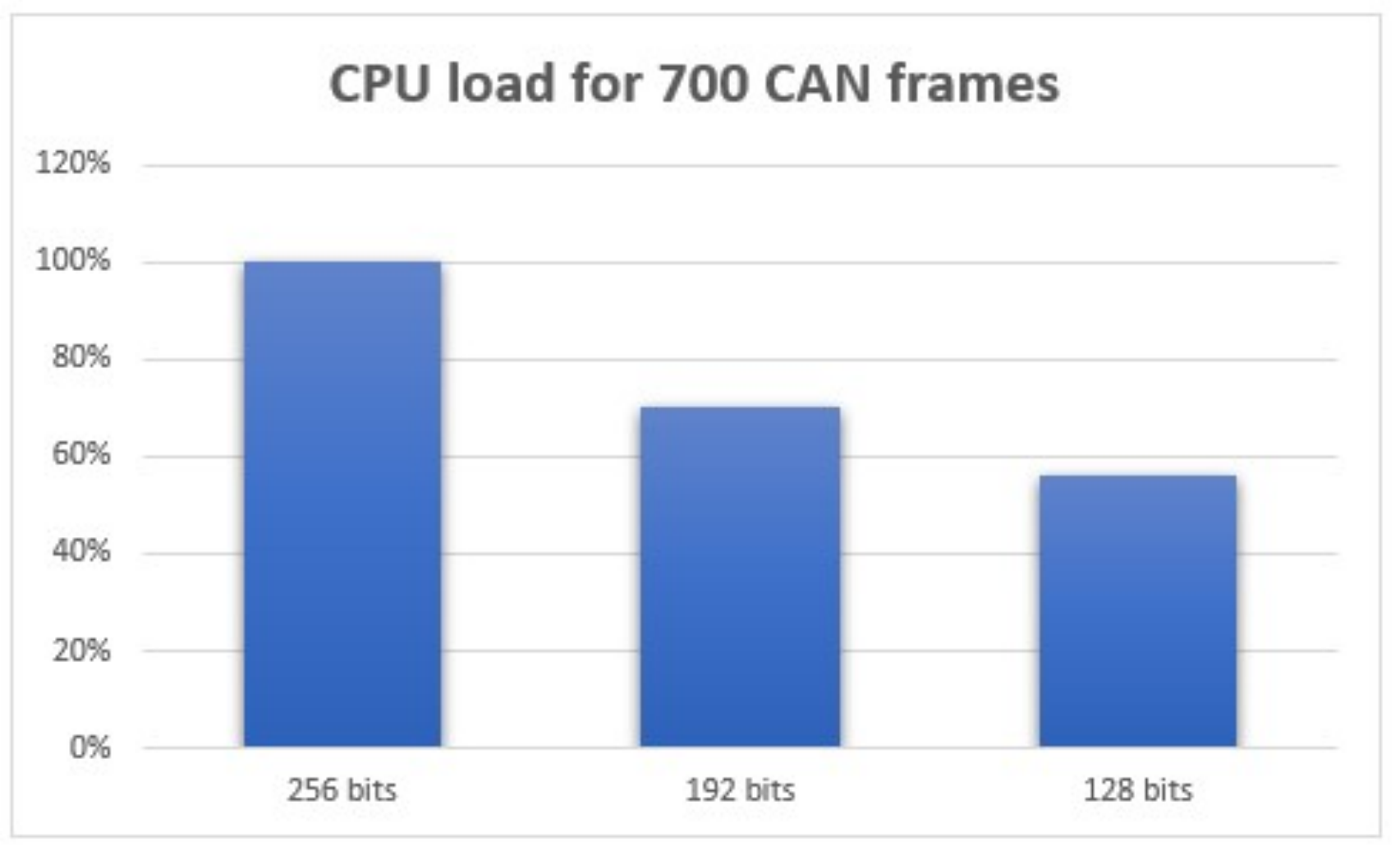}
         \caption{\scriptsize{CPU utilization trend keeping constant the number of processed frames for different CMAC digest's lengths.}}
         \label{fig:cpuvar}
     \end{subfigure}
    \label{fig:TAURUMdata}
    \caption{CPU execution time saving in shorter key mode}
\end{figure}

As discussed in \autoref{sec:keyprov}, EXT-TAURUM P2T introduces a Short  Secret  Key mode that is set at run-time in case of need. The following experimental results are focused on determining the throughput overhead trend introduced by CMAC calculation with different key lengths. 
\autoref{fig:maxnframes} shows the maximum number of CMAC digest computations that the system can sustain in the hypothesis of dedicating all resources to this activity. At the same time, \autoref{fig:cpuvar}  shows the saving in terms of resources changing CMAC digest data length. The figure clearly shows how reducing the CMAC digest from 256 bit to 128bit enables about 40\% saving of CPU time that can be used to handle critical overloading situations.

\begin{figure}[ht!]
\centering
\includegraphics[width=\columnwidth]{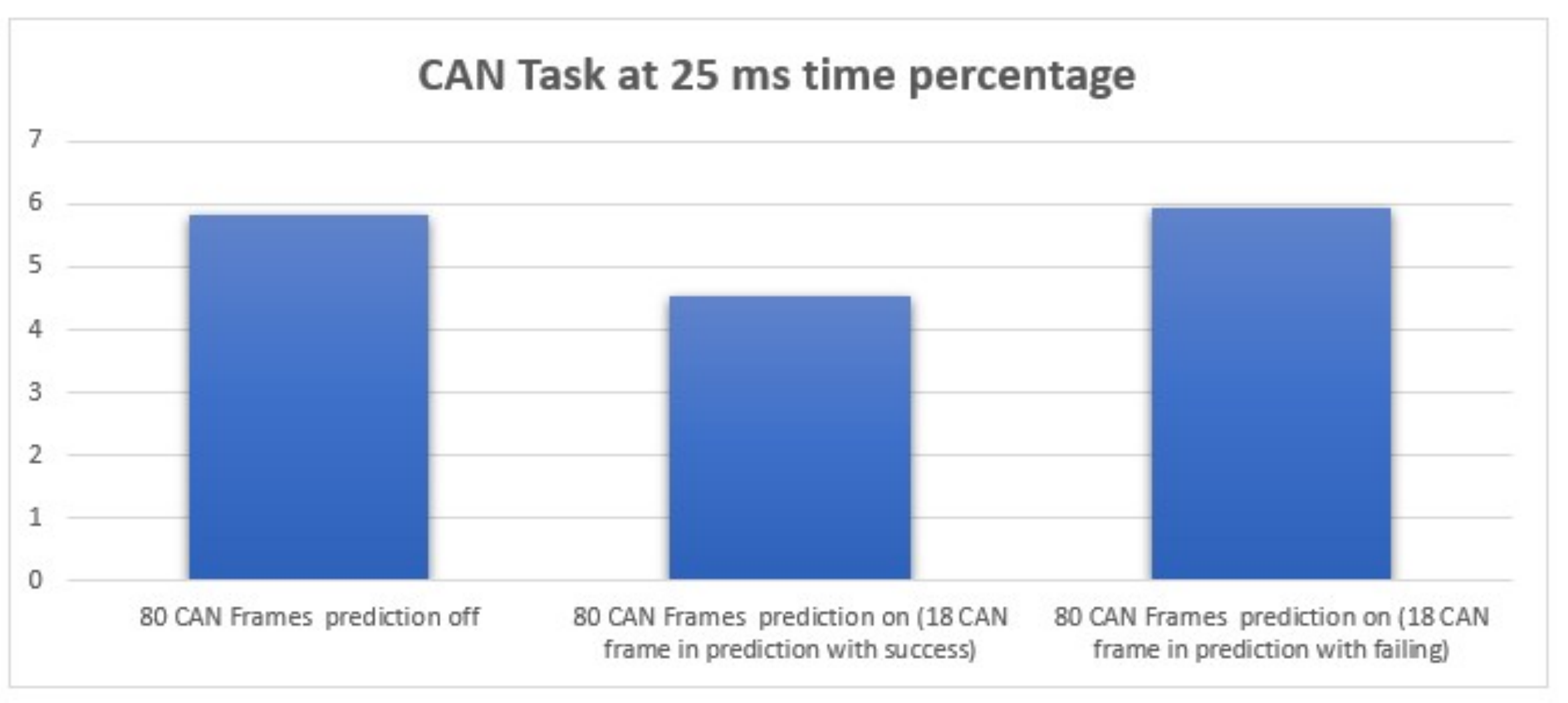}
\caption{TAURUM P2T Advanced Secure CAN network throughput trend profile based on HMAC used mode}
\label{fig:tp}
\end{figure}

\autoref{fig:tp} reports results concerning the speculative MAC calculation, described in \autoref{sec:speMac}.
The target reference for this experiment is a periodic task scheduled every 25ms and processing 80 different CAN frames whose CMAC digests must be authenticated. 18 out of the 80 processed frames transmit steady-state information, and their authentication can benefit from speculative MAC calculation. 

With speculative MAC disabled, the frame authentication requires around 6\% of real-time CPU time in regular running. 

By activating speculative MAC calculation on the 18 steady-state frames, in the hypothesis that all speculations are successful, the real-time CPU usage drops down to around 1\%, demonstrating the effectiveness of this technique.

Finally, in the worst-case condition of all speculations failing, a 0.1\% real-time CPU usage overhead is introduced, with a negligible impact on the system.

\subsection{Overhead evaluation}

Implementing the EXT-TAURUM P2T communication stack introduces extra code. Comparing the firmware of one of our sample nodes implemented without any security feature with one implementing the EXT-TAURUM P2T communication stack, we measured a 300\% code overhead. Nevertheless, in our prototype, all cryptographic operations are software implemented. In real ECUs, the use of Crypto Cores would significantly mitigate this overhead.  
As described before, at the very first time, the system executes the key provisioning protocol for sharing and exchanging keys to all the ECUs in the network. This process lasts no more than 50 ms for sharing the secret keys between a secure gateway and two secure nodes in our experimental implementation. This time strongly depends on the CAN baud-rate settings. A similar amount of time, less than 50ms, is also needed to update secret keys for each privilege level at the end of each rolling period. Being this a broadcast operation, this time is not influenced by the number of nodes. Time measurements are all performed on the prototype implementation of the system.  
Experimental activities also proved the concept's capability on privilege separation.

Eventually, let us discuss the EXT-TAURUM P2T impact on the hardware architecture. Most of the ECUs in actual vehicles are already multi-CAN devices, often with spare channels available. They, therefore, are already able to host two CAN-buses. Thus, the EXT-TAURUM P2T architecture only requires adding the SGTW module and ensuring the Secure CAN cabling. This hardware overhead is mitigated by the lack of the IT secret key management infrastructure, with annexed security weakness described above and impact on management costs.

\subsection{Security analysis}

The proof of the security of the EXT-TAURUM P2T solution holds under the {\em infeasibility} hypothesis. We assume the use of state-of-the-art secure cryptographic algorithms with proper key lengths. 

The keystone of EXT-TAURUM P2T is a mechanism to establish a first shared secret $K_{SH}$representing a root of trust for all following security mechanisms. EXT-TAURUM P2T exploits state-of-the-art Elliptic curve cryptography (ECC), a public-key cryptography schema suitable for use in environments with limited resources such as mobile devices and smart cards. In particular, EXT-TAURUM P2T exploits Curve25519, an elliptic curve that offers state-of-the-art 128 security bits and is designed for use in the Elliptic Curve Diffie-Hellman (ECDH) key agreement design scheme. This curve is one of the fastest ECC curves and more resistant to the weak number random generator. Curve25519 is built in such a way as to avoid potential attacks on implementation and avoid side-channel attacks and random number generator issues.

After establishing the first shared secret, all communications on the secure CAN network are encrypted using Advanced Encryption Standard (AES), implemented with the state-of-the-art AES256 Cipher with Block Chaining (CBC) modality. This guarantees confidentiality, integrity, and authentication of all messages transmitted over this channel. 

Regarding messages exchanged over the public CAN network, integrity and authenticity are implemented exploiting state-of-the-art Cipher-based Message Authentication Code (CMAC). Confidentiality on this network cannot be introduced since legislation requirements impose plaintext transmission on this network. MAC exploits secret keys securely shared among parties using the secure CAN network. A rolling counter mechanism is used to avoid reply attacks. The introduction of the privilege level concept, EXT-TAURUM P2T, compartmentalizes the security level (i.e., CMAC key length) implemented on this channel. State-of-the-art AES256 is used at the higher levels, while lower levels can resort to reduced key lengths. Even if reduced key lengths might represent a security threat, implementing the periodic key rolling protocol guarantees a minimal timeframe to mount an attack. The key deprecation protocol ensures a secure approach to react if a secret is compromised.

MitM attacks on the network can be efficiently prevented with the above mechanisms. Moreover, the availability of a secure communication channel enables secure authentication of each hardware module in the CAN network resorting to the protocol provided in Section IV-C. According to this protocol, every ECU connected to the network embeds a secret provided by the carmaker at the plant. The protocol exploits authentication using the nonce to identify trusted modules and resorts to the security of the secure CAN network to accomplish the required exchange of messages.

All previous security mechanisms require state-of-the-art hardware blocks to securely store secret keys and perform cryptographic operations onboard each ECU connected to the CAN network. However, this is a standard requirement in the automotive domain where ECUs are equipped with dedicated Hardware Security Modules. The basic assumption is that these modules are secure against costly physical attacks such as side-channel attacks. Moreover, thanks to the key provisioning protocol introduced by EXT TAURUM-P2T, even if an attacker succeeds in performing a physical attack able to compromise a single vehicle, the effect of the attack will be limited in time to the key rolling period and limited in space to a single vehicle and not to the entire fleet.

To conclude the security analyses, EXT-TAURUM P2T can mitigate DoS attacks even if it cannot altogether remove this threat. Mitigation is introduced by introducing reduced key lengths for cryptographic operations and speculative MAC computation. Both solutions can be exploited to reduce the system's load whenever computation and transmission peaks typical of DoS attacks arise in the system.

\section{Conclusion}
\label{sec:conclusions}


The ever-increasing adoption of electronic-based systems in road vehicles has opened the door for new security vulnerabilities in modern designs. EXT-TAURUM P2T Advanced Secure CAN-FD Architecture protects the vehicle's communication infrastructure by implementing new security features including: (i) a periodic secure key provisioning mechanism that exploits the architecture's secure channel, (ii) the implementation of privilege levels of security by separating trust zones from untrusted ones and (iii) the dynamic reallocation of the MAC computations to a background task that reduced the utilization of the CPU for real-time computations.
The new features proposed in this paper have been experimentally validated through a set of experiments whose results assess its feasibility. 
Finally, a preliminary cost evaluation of a possible industrial implementation of the proposed architecture shows that the proposed EXT-TAURUM P2T can be affordably produced.


%





\ifCLASSOPTIONcaptionsoff
  \newpage
\fi



%
%
\nocite{*}
\bibliographystyle{./bibliography/IEEEtran}
\bibliography{./bibliography/IEEEabrv,./bibliography/TP2T}

\begin{thebibliography}{10}
\providecommand{\url}[1]{#1}
\csname url@samestyle\endcsname
\providecommand{\newblock}{\relax}
\providecommand{\bibinfo}[2]{#2}
\providecommand{\BIBentrySTDinterwordspacing}{\spaceskip=0pt\relax}
\providecommand{\BIBentryALTinterwordstretchfactor}{4}
\providecommand{\BIBentryALTinterwordspacing}{\spaceskip=\fontdimen2\font plus
\BIBentryALTinterwordstretchfactor\fontdimen3\font minus
  \fontdimen4\font\relax}
\providecommand{\BIBforeignlanguage}[2]{{%
\expandafter\ifx\csname l@#1\endcsname\relax
\typeout{** WARNING: IEEEtran.bst: No hyphenation pattern has been}%
\typeout{** loaded for the language `#1'. Using the pattern for}%
\typeout{** the default language instead.}%
\else
\language=\csname l@#1\endcsname
\fi
#2}}
\providecommand{\BIBdecl}{\relax}
\BIBdecl

\bibitem{WP29}
\BIBentryALTinterwordspacing
{UN Economic Commission for Europe}, ``Unece world forum for harmonization of
  vehicle regulations (wp.29),'' 2021. [Online]. Available:
  \url{https://unece.org/wp29-introduction}
\BIBentrySTDinterwordspacing

\bibitem{unece-155-2021}
\BIBentryALTinterwordspacing
------, ``Un regulation no. 155 - cyber security and cyber security management
  system,'' 2021. [Online]. Available:
  \url{https://unece.org/transport/documents/2021/03/standards/un-regulation-no-155-cyber-security-and-cyber-security}
\BIBentrySTDinterwordspacing

\bibitem{unece-156-2021}
\BIBentryALTinterwordspacing
------, ``Un regulation no. 156 - software update and software update
  management system,'' 2021. [Online]. Available:
  \url{https://unece.org/transport/documents/2021/03/standards/un-regulation-no-156-software-update-and-software-update}
\BIBentrySTDinterwordspacing

\bibitem{bozdal2020evaluation}
M.~Bozdal, M.~Samie, S.~Aslam, and I.~Jennions, ``Evaluation of can bus
  security challenges,'' \emph{Sensors}, vol.~20, no.~8, p. 2364, 2020.

\bibitem{6884472}
T.~{Nguyen}, B.~M. {Cheon}, and J.~W. {Jeon}, ``Can fd performance analysis for
  ecu re-programming using the canoe,'' in \emph{The 18th IEEE International
  Symposium on Consumer Electronics (ISCE 2014)}, 2014, pp. 1--4.

\bibitem{oberti2021taurum}
F.~Oberti, E.~Sanchez, A.~Savino, F.~Parisi, and S.~Di~Carlo, ``Taurum p2t:
  Advanced secure can-fd architecture for road vehicle,'' in \emph{2021 IEEE
  27th International Symposium on On-Line Testing and Robust System Design
  (IOLTS)}.\hskip 1em plus 0.5em minus 0.4em\relax IEEE, 2021, pp. 1--7.

\bibitem{9525579}
------, ``Mitigation of automotive control modules hardware replacement-based
  attacks through hardware signature,'' in \emph{2021 51st Annual IEEE/IFIP
  International Conference on Dependable Systems and Networks - Supplemental
  Volume (DSN-S)}, 2021, pp. 13--14.

\bibitem{albert2004comparison}
A.~Albert \emph{et~al.}, ``Comparison of event-triggered and time-triggered
  concepts with regard to distributed control systems,'' \emph{Embedded world},
  vol. 2004, pp. 235--252, 2004.

\bibitem{CIAtriad}
M.~U. Farooq, M.~Waseem, A.~Khairi, and S.~Mazhar, ``A critical analysis on the
  security concerns of internet of things (iot),'' \emph{International Journal
  of Computer Applications}, vol. 111, no.~7, 2015.

\bibitem{CMAC}
\BIBentryALTinterwordspacing
{Network Working Group}, ``The aes-cmac algorithm,'' 2021. [Online]. Available:
  \url{https://tools.ietf.org/html/rfc4493.html}
\BIBentrySTDinterwordspacing

\bibitem{MAC}
N.~{Nowdehi}, A.~{Lautenbach}, and T.~{Olovsson}, ``In-vehicle can message
  authentication: An evaluation based on industrial criteria,'' in \emph{2017
  IEEE 86th Vehicular Technology Conference (VTC-Fall)}, 2017, pp. 1--7.

\bibitem{Replyattack}
M.~{Marchetti} and D.~{Stabili}, ``Anomaly detection of can bus messages
  through analysis of id sequences,'' in \emph{2017 IEEE Intelligent Vehicles
  Symposium (IV)}, 2017, pp. 1577--1583.

\bibitem{rollingcounter}
Y.~Xiao, H.-H. Chen, R.~Wang, and S.~Sethi, ``Mac security and security
  overhead analysis in the ieee 802.15.4 wireless sensor networks,''
  \emph{EURASIP Journal on Wireless Communications and Networking}, vol. 2006,
  04 2006.

\bibitem{osek}
\BIBentryALTinterwordspacing
{ISO - International Organization for Standardization}, ``Iso 17356-2 road
  vehicles --- open interface for embedded automotive applications --- part 2:
  Osek/vdx specifications for binding os, com and nm,'' 2005. [Online].
  Available: \url{https://www.iso.org/standard/33007.html}
\BIBentrySTDinterwordspacing

\bibitem{kluge2009implementing}
F.~Kluge, C.~Yu, J.~Mische, S.~Uhrig, and T.~Ungerer, ``Implementing autosar
  scheduling and resource management on an embedded smt processor,'' in
  \emph{Proceedings of th 12th International Workshop on Software and Compilers
  for Embedded Systems}, 2009, pp. 33--42.

\bibitem{8762043}
M.~S.~U. {Alam}, S.~{Iqbal}, M.~{Zulkernine}, and C.~{Liem}, ``Securing vehicle
  ecu communications and stored data,'' in \emph{ICC 2019 - 2019 IEEE
  International Conference on Communications (ICC)}, 2019, pp. 1--6.

\bibitem{8474730}
R.~{R.V.} and K.~{A.}, ``Secure boot of embedded applications - a review,'' in
  \emph{2018 Second International Conference on Electronics, Communication and
  Aerospace Technology (ICECA)}, 2018, pp. 291--298.

\bibitem{6542519}
C.~{Lin} and A.~{Sangiovanni-Vincentelli}, ``Cyber-security for the controller
  area network (can) communication protocol,'' in \emph{2012 International
  Conference on Cyber Security}, 2012, pp. 1--7.

\bibitem{6379926}
H.~{Kang}, Y.~{Hori}, and A.~{Satoh}, ``Performance evaluation of the first
  commercial puf-embedded rfid,'' in \emph{The 1st IEEE Global Conference on
  Consumer Electronics 2012}, 2012, pp. 5--8.

\bibitem{9291788}
R.~{Soga} and H.~{Kang}, ``Physical unclonable function using carbon
  resistor,'' in \emph{2020 IEEE 9th Global Conference on Consumer Electronics
  (GCCE)}, 2020, pp. 559--561.

\bibitem{7362173}
M.~{Yasin}, J.~J. {Rajendran}, O.~{Sinanoglu}, and R.~{Karri}, ``On improving
  the security of logic locking,'' \emph{IEEE Transactions on Computer-Aided
  Design of Integrated Circuits and Systems}, vol.~35, no.~9, pp. 1411--1424,
  2016.

\bibitem{9070188}
K.~{Juretus} and I.~{Savidis}, ``Increased output corruption and structural
  attack resilience for sat attack secure logic locking,'' \emph{IEEE
  Transactions on Computer-Aided Design of Integrated Circuits and Systems},
  vol.~40, no.~1, pp. 38--51, 2021.

\bibitem{8192439}
T.~{Thangam}, G.~{Gayathri}, and T.~{Madhubala}, ``A novel logic locking
  technique for hardware security,'' in \emph{2017 IEEE International
  Conference on Electrical, Instrumentation and Communication Engineering
  (ICEICE)}, 2017, pp. 1--7.

\bibitem{AES}
\BIBentryALTinterwordspacing
{Network Working Group}, ``The aes-cbc cipher algorithm and its use with
  ipsec,'' 2021. [Online]. Available: \url{https://tools.ietf.org/html/rfc3602}
\BIBentrySTDinterwordspacing

\bibitem{Koblitz04guideto}
N.~Koblitz, A.~Menezes, and S.~Vanstone, ``Guide to elliptic curve
  cryptography,'' 2004.

\bibitem{10.1007/11745853_14}
D.~J. Bernstein, ``Curve25519: New diffie-hellman speed records,'' in
  \emph{Public Key Cryptography - PKC 2006}, M.~Yung, Y.~Dodis, A.~Kiayias, and
  T.~Malkin, Eds.\hskip 1em plus 0.5em minus 0.4em\relax Berlin, Heidelberg:
  Springer Berlin Heidelberg, 2006, pp. 207--228.

\bibitem{NEOVI}
\BIBentryALTinterwordspacing
{Intrepid Control Systems, Inc}, ``neovi fire 2 user guide,'' 2021. [Online].
  Available:
  \url{URL:https://cdn.intrepidcs.net/guides/neovifire2/neovi_fire2_ug.pdf}
\BIBentrySTDinterwordspacing

\bibitem{inbook}
G.~Macher, C.~Schmittner, O.~Veledar, and E.~Brenner, \emph{ISO/SAE DIS 21434
  Automotive Cybersecurity Standard - In a Nutshell}, 09 2020, pp. 123--135.

\bibitem{7092463}
G.~{Macher}, H.~{Sporer}, R.~{Berlach}, E.~{Armengaud}, and C.~{Kreiner},
  ``Sahara: A security-aware hazard and risk analysis method,'' in \emph{2015
  Design, Automation Test in Europe Conference Exhibition (DATE)}, 2015, pp.
  621--624.

\bibitem{8587713}
R.~G. {Dutta}, F.~{Yu}, T.~{Zhang}, Y.~{Hu}, and Y.~{Jin}, ``Security for
  safety: A path toward building trusted autonomous vehicles,'' in \emph{2018
  IEEE/ACM International Conference on Computer-Aided Design (ICCAD)}, 2018,
  pp. 1--6.

\bibitem{bosch}
F.~Hartwich and R.~P. Bosch, ``Can with flexible data-rate,'' 2012.

\bibitem{7457584}
I.~S. for Information~technology ISO, ``Iso/iec/ieee international standard for
  information technology -- telecommunications and information exchange between
  systems -- local and metropolitan area networks -- part 1ae: Media access
  control (mac) security - amendment 1: Galois counter model -- advanced
  encryption standard-256 (gcmaes-256) cipher suite,'' \emph{ISO/IEC/ IEEE
  8802-1AE First edition 2013-12-01 AMENDMENT 1 2015-05-01}, pp. 1--57, 2015.

\bibitem{7579937}
K.~{Kang}, Y.~{Baek}, S.~{Lee}, and S.~H. {Son}, ``Lightweight authentication
  method for controller area network,'' in \emph{2016 IEEE 22nd International
  Conference on Embedded and Real-Time Computing Systems and Applications
  (RTCSA)}, 2016, pp. 101--101.

\bibitem{7435304}
S.~{Woo}, H.~J. {Jo}, I.~S. {Kim}, and D.~H. {Lee}, ``A practical security
  architecture for in-vehicle can-fd,'' \emph{IEEE Transactions on Intelligent
  Transportation Systems}, vol.~17, no.~8, pp. 2248--2261, 2016.

\bibitem{6409452}
B.~{Groza} and S.~{Murvay}, ``Efficient protocols for secure broadcast in
  controller area networks,'' \emph{IEEE Transactions on Industrial
  Informatics}, vol.~9, no.~4, pp. 2034--2042, 2013.

\bibitem{6001769}
R.~I. {Davis}, S.~{Kollmann}, V.~{Pollex}, and F.~{Slomka}, ``Controller area
  network (can) schedulability analysis with fifo queues,'' in \emph{2011 23rd
  Euromicro Conference on Real-Time Systems}, 2011, pp. 45--56.

\bibitem{6730667}
P.~{Murvay} and B.~{Groza}, ``Source identification using signal
  characteristics in controller area networks,'' \emph{IEEE Signal Processing
  Letters}, vol.~21, no.~4, pp. 395--399, 2014.

\bibitem{6471884}
H.~{Nicanfar} and V.~C.~M. {Leung}, ``Multilayer consensus ecc-based password
  authenticated key-exchange (mcepak) protocol for smart grid system,''
  \emph{IEEE Transactions on Smart Grid}, vol.~4, no.~1, pp. 253--264, 2013.

\bibitem{7182505}
D.~{Jang}, S.~{Han}, S.~{Kang}, and J.~{Choi}, ``Communication channel modeling
  of controller area network (can),'' in \emph{2015 Seventh International
  Conference on Ubiquitous and Future Networks}, 2015, pp. 86--88.

\bibitem{5116731}
H.~{Chen} and J.~{Tian}, ``Research on the controller area network,'' in
  \emph{2009 International Conference on Networking and Digital Society},
  vol.~2, 2009, pp. 251--254.

\bibitem{8053478}
J.~{Liu}, S.~{Zhang}, W.~{Sun}, and Y.~{Shi}, ``In-vehicle network attacks and
  countermeasures: Challenges and future directions,'' \emph{IEEE Network},
  vol.~31, no.~5, pp. 50--58, 2017.

\bibitem{7995934}
M.~{Marchetti} and D.~{Stabili}, ``Anomaly detection of can bus messages
  through analysis of id sequences,'' in \emph{2017 IEEE Intelligent Vehicles
  Symposium (IV)}, 2017, pp. 1577--1583.

\bibitem{8004161}
Y.~{Zhang}, M.~{Chen}, N.~{Guizani}, D.~{Wu}, and V.~C.~M. {Leung}, ``Sovcan:
  Safety-oriented vehicular controller area network,'' \emph{IEEE
  Communications Magazine}, vol.~55, no.~8, pp. 94--99, 2017.

\bibitem{5340585}
M.~{Barranco}, J.~{Proenza}, and L.~{Almeida}, ``Quantitative comparison of the
  error-containment capabilities of a bus and a star topology in can
  networks,'' \emph{IEEE Transactions on Industrial Electronics}, vol.~58,
  no.~3, pp. 802--813, 2011.

\bibitem{5582170}
P.~{Mart{\'\i}}, A.~{Camacho}, M.~{Velasco}, and M.~E.~M. {Ben Gaid}, ``Runtime
  allocation of optional control jobs to a set of can-based networked control
  systems,'' \emph{IEEE Transactions on Industrial Informatics}, vol.~6, no.~4,
  pp. 503--520, 2010.

\bibitem{8303766}
W.~{Choi}, H.~J. {Jo}, S.~{Woo}, J.~Y. {Chun}, J.~{Park}, and D.~H. {Lee},
  ``Identifying ecus using inimitable characteristics of signals in controller
  area networks,'' \emph{IEEE Transactions on Vehicular Technology}, vol.~67,
  no.~6, pp. 4757--4770, 2018.

\bibitem{6242539}
P.~M. {Yomsi}, D.~{Bertrand}, N.~{Navet}, and R.~I. {Davis}, ``Controller area
  network (can): Response time analysis with offsets,'' in \emph{2012 9th IEEE
  International Workshop on Factory Communication Systems}, 2012, pp. 43--52.

\bibitem{6891169}
C.~{Lin}, Q.~{Zhu}, and A.~{Sangiovanni-Vincentelli}, ``Security-aware modeling
  and efficient mapping for can-based real-time distributed automotive
  systems,'' \emph{IEEE Embedded Systems Letters}, vol.~7, no.~1, pp. 11--14,
  2015.

\bibitem{8457262}
B.~{Groza} and P.~{Murvay}, ``Efficient intrusion detection with bloom
  filtering in controller area networks,'' \emph{IEEE Transactions on
  Information Forensics and Security}, vol.~14, no.~4, pp. 1037--1051, 2019.

\bibitem{8946711}
P.~{Nuzzo}, N.~{Bajaj}, M.~{Masin}, D.~{Kirov}, R.~{Passerone}, and A.~L.
  {Sangiovanni-Vincentelli}, ``Optimized selection of reliable and
  cost-effective safety-critical system architectures,'' \emph{IEEE
  Transactions on Computer-Aided Design of Integrated Circuits and Systems},
  vol.~39, no.~10, pp. 2109--2123, 2020.

\bibitem{KOPPERMANN2017491}
\BIBentryALTinterwordspacing
P.~Koppermann, F.~{De Santis}, J.~Heyszl, and G.~Sigl, ``Low-latency x25519
  hardware implementation: breaking the 100 microseconds barrier,''
  \emph{Microprocessors and Microsystems}, vol.~52, pp. 491--497, 2017.
  [Online]. Available:
  \url{https://www.sciencedirect.com/science/article/pii/S0141933117300273}
\BIBentrySTDinterwordspacing

\bibitem{6855562}
F.~{Stellari}, P.~{Song}, A.~J. {Weger}, J.~{Culp}, A.~{Herbert}, and
  D.~{Pfeiffer}, ``Verification of untrusted chips using trusted layout and
  emission measurements,'' in \emph{2014 IEEE International Symposium on
  Hardware-Oriented Security and Trust (HOST)}, 2014, pp. 19--24.

\bibitem{5737786}
M.~{Majzoobi} and F.~{Koushanfar}, ``Time-bounded authentication of fpgas,''
  \emph{IEEE Transactions on Information Forensics and Security}, vol.~6,
  no.~3, pp. 1123--1135, 2011.

\bibitem{6993402}
M.~B. {Bahador}, M.~{Abadi}, and A.~{Tajoddin}, ``Hpcmalhunter: Behavioral
  malware detection using hardware performance counters and singular value
  decomposition,'' in \emph{2014 4th International Conference on Computer and
  Knowledge Engineering (ICCKE)}, 2014, pp. 703--708.

\bibitem{SPEC}
H.~Aydin, R.~Melhem, D.~Mosse, and P.~Mejia-Alvarez, ``Power-aware scheduling
  for periodic real-time tasks,'' \emph{IEEE Transactions on Computers},
  vol.~53, no.~5, pp. 584--600, 2004.

\bibitem{prediction}
K.~Kinkai, T.~Baba, H.~Jutori, K.~Ootsu, T.~Ohkawa, and T.~Yokota,
  ``Comparative study of path prediction method for speculative loop
  execution,'' in \emph{2012 Third International Conference on Networking and
  Computing}, 2012, pp. 283--287.

\end{thebibliography}
%

\begin{IEEEbiography}
[{\includegraphics[width=1in,height=1.25in,clip,keepaspectratio]{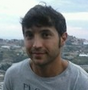}}]
{Franco Oberti}
Franco Oberti (student, IEEE '20                                                                                                                                                                                                                                                                                                                                                                                                                                                                                                                                                                                                                                                                                                                                                                                                                                                                                                                                                                                                                                                                                                                  ) received the M.Sc. degrees in computer engineering from the Politecnico di Torino, Torino, Italy, in 2007. it started working in PUNCH Torino (former General Motor Powertrain Europe) in 2007, where he held different positions. In 2016 he received a Master certificate from Standford University in Advanced Cybersecurity. Currently, he is part of the Product Security Office in PUNC Torino. By 2021, he was also an Industry PhD student candidate. His current research interests include cybersecurity applied to an embedded system in the road vehicles domain.
\end{IEEEbiography}

\begin{IEEEbiography}[{\includegraphics[width=1in,height=1.25in,clip,keepaspectratio]{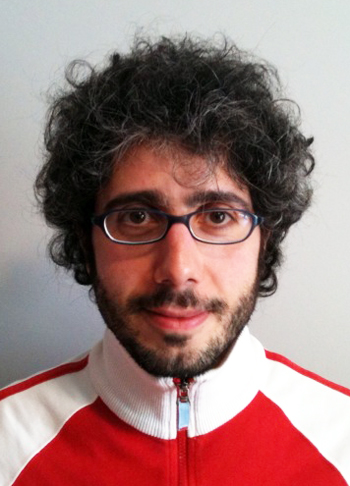}}]{Alessandro Savino}(M'14) is an Assistant Professor in the Department of Control and Computer Engineering at Politecnico di Torino (Italy). He holds a Ph.D. (2009) and an M.S. equivalent (2005) in Computer Engineering and Information Technology from the Politecnico di Torino in Italy. Dr. Savino's research contributions include Approximate Computing, Reliability Analysis, Safety-Critical Systems, Software-Based Self-Test, and Image Analysis. He has been part of the program and organizing committee of several IEEE and INSTICC conferences and served as a reviewer of IEEE conferences and journals. His research interests include Operating Systems, Imaging algorithms, Machine Learning, Evolutionary Algorithms, Graphical User Interface experience, and Audio manipulation.
\end{IEEEbiography}


\begin{IEEEbiography}[{\includegraphics[width=1in,height=1.25in,clip,keepaspectratio]{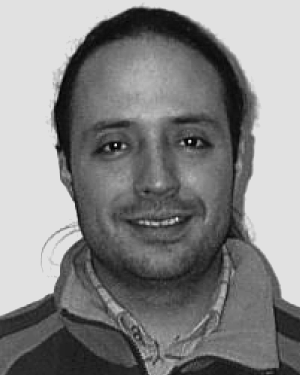}}]{Ernesto Sanchez}
received the degree in electronic engineering from Universidad Javeriana, Bogota, Colombia, in 2000, and the Ph.D. degree in computer engineering from the Politecnico di Torino, Italy, in 2006, where he is currently an Associate Professor in the Department of Control and Computer Engineering. His main research interests include microprocessor testing, hardware security and DNN reliability.

\end{IEEEbiography}

\begin{IEEEbiography}[{\includegraphics[width=1in,height=1.25in,clip,keepaspectratio]{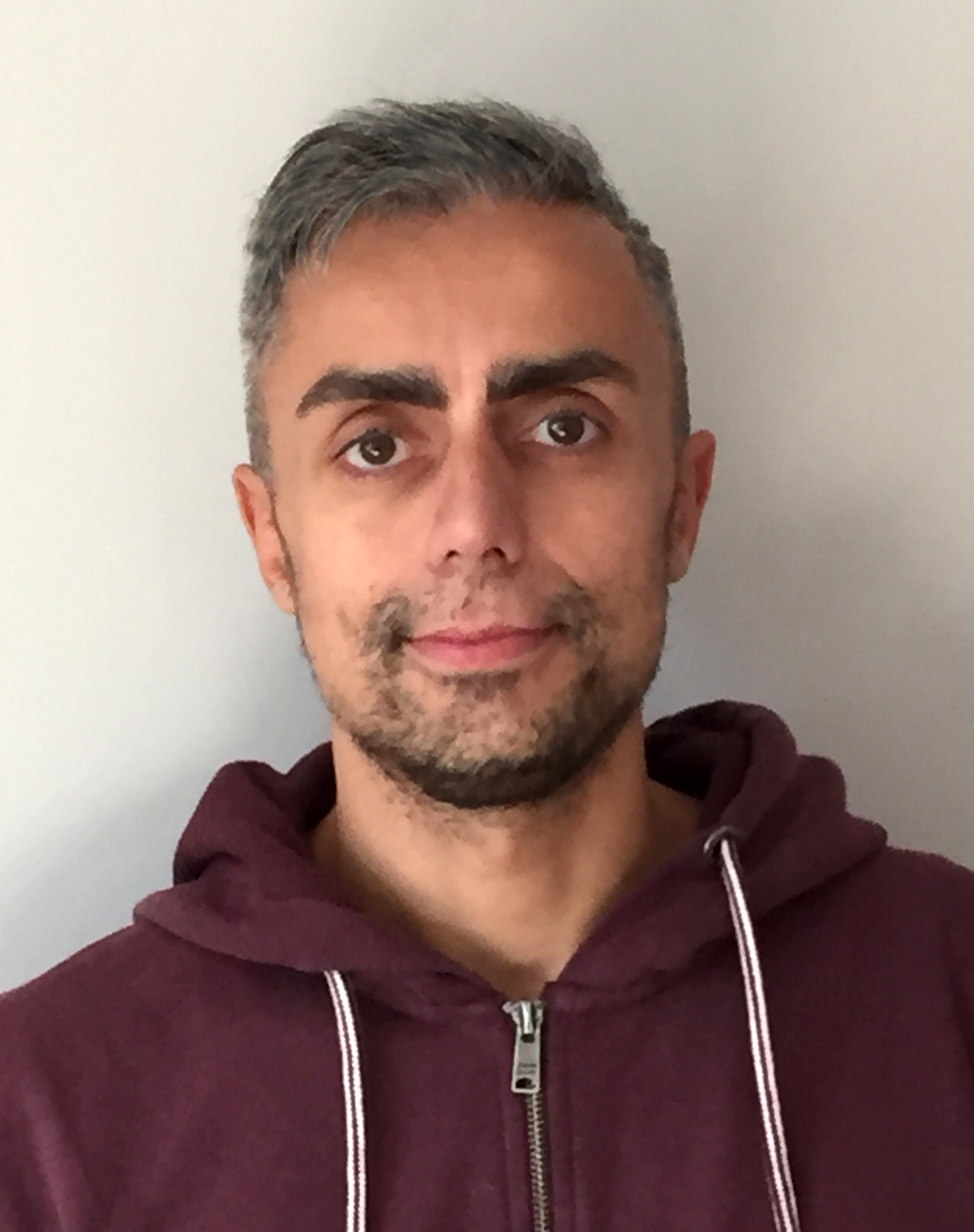}}]{Stefano Di Carlo}
(SM'00-M'03-SM'11) received a M.Sc. degree in computer engineering and a Ph.D. degree in information technologies from Politecnico di Torino, Italy, where he is a tenured Associate professor. His research interests include DFT, BIST, and dependability. He has coordinated several national and European research projects. Di Carlo has published more than 200 papers in peer reviewed IEEE and ACM journals and conferences. He regularly serves on the Organizing and Program Committees of major IEEE and ACM conferences. He is a golden core member of the IEEE Computer Society and a senior member of the IEEE.
\end{IEEEbiography}

\begin{IEEEbiography}
[{\includegraphics[width=1in,height=1.25in,clip,keepaspectratio]{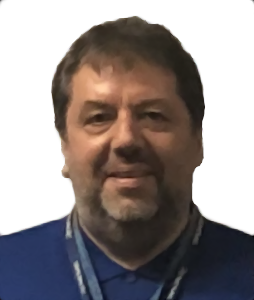}}]
{Filippo Parisi}
Filippo Parisi, aka albix, received a degree in electronic engineering from Politecnico of Turin, Turin, Italy, in 1992. As manager in PUNCH Torino, he is leading the development of electronics, firmware and virtualization for testing applied to hard real-time, safety-critical automotive embedded control systems. He held several positions in multinational automotive companies as FIAT Research Center, FIAT-GM-Powertrain JV and General Motors for more than 25 years.
\end{IEEEbiography}



\end{document}